\documentclass[twocolumn,10pt]{asme2ej}
\usepackage{graphicx} 
\usepackage{hyperref} 
\usepackage[shortlabels]{enumitem}
\setlist[itemize]{label=\textbullet}
\hypersetup{
	colorlinks=true,
	linkcolor=blue,
	citecolor=blue,
	urlcolor=blue,
}
\usepackage[utf8]{inputenc}
\usepackage{lipsum}
\usepackage[square,numbers]{natbib}
\usepackage{xcolor,soul,framed} 
\colorlet{shadecolor}{yellow}
\usepackage[cmex10]{amsmath}
\usepackage{array}
\usepackage{mdwmath}
\usepackage{mdwtab}
\usepackage{eqparbox}
\usepackage{url}
\usepackage{MnSymbol}%
\usepackage{wasysym}%
\newtheorem{theorem}{Theorem}
\newtheorem{lemma}{Lemma}

\newtheorem{proposition}{Proposition}
\newtheorem{definition}[theorem]{Definition}
\newtheorem{remark}[theorem]{Remark}

\title{On the Contraction Analysis of Nonlinear System with Multiple Equilibrium Points}

\author{Riddhi Mohan Bora*
    \affiliation{
	Student Member, IEEE \\
	Department of Electrical Engineering\\
	Indian Institute of Technology Delhi\\
	Hauz Khas, New Delhi, India, 110016\\
    Email: Riddhi.Mohan.Bora@iitd.ac.in
    }	
}

\author{Bhabani Shankar Dey 
    \affiliation{ Student Member, IEEE  \\
    	Department of Electrical Engineering\\
	Indian Institute of Technology Delhi\\
	Hauz Khas, New Delhi, India, 110016\\
    Email: Bhabani.Shankar.Dey@iitd.ac.in
    }	
}

\author{Indra Narayan Kar 
    \affiliation{Senior Member, IEEE  \\
    Professor, Department of Electrical Engineering\\
	Indian Institute of Technology Delhi\\
	Hauz Khas, New Delhi, India, 110016\\
    Email: ink@ee.iitd.ac.in
    }	
}
\begin{document}
\maketitle    
\begin{abstract}
{\it In this work, we leverage the 2-contraction theory, which extends the capabilities of classical contraction theory, to develop a global stability framework. Coupled with powerful geometric tools such as Poincaré index theory, the 2-contraction theory enables us to analyze the stability of planar nonlinear systems without relying on local equilibrium analysis. By utilizing index theory and 2-contraction results, we efficiently characterize the nature of equilibrium points and delineate regions in $\Re^2$ where periodic solutions, closed orbits, or stable dynamics may exist. A key focus of this work is the identification of regions in the state space where periodic solutions may occur, as well as 2-contraction regions that guarantee the nonexistence of such solutions. Additionally, we address a critical problem in engineering—the determination of the basin of attraction (BOA) for stable equilibrium points. For systems with multiple equilibria identifying candidate BOAs becomes highly nontrivial. We propose a novel methodology leveraging the 2-contraction theory to approximate a common BOA for a class of nonlinear systems with multiple stable equilibria. Theoretical findings are substantiated through benchmark examples and numerical simulations, demonstrating the practical utility of the proposed approach. Furthermore, we extend our framework to analyze networked systems, showcasing their efficacy in an opinion dynamics problem. 
}
\end{abstract}
\section{Introduction}
Stability has been a cardinal aspect of control system design and analysis. When it comes to the stability analysis of nonlinear systems, the notion of equilibrium state or nominal motion comes into the picture. If the nonlinear system is equipped with a unique equilibrium point, commenting on global stability around that has been studied extensively in the literature. On the contrary, numerous systems have multiple equilibrium points. Commenting on the overall stability in such cases has been sparsely solved in literature. Conventional stability notions comment on the stability around an equilibrium point. In the case of multiple equilibria, the stability analysis has revolved around the respective points of interest. As in existing literature, we often encounter many nonlinear dynamical system models, such as opinion dynamics \cite{xiong2018social, proskurnikov2015opinion, doi:10.1137/130913250}, epidemiology model \cite{hesaaraki2013global,intissar2020application} and Lurie problem \cite{RON_OFIR,9782543}, etc. which exhibit multiple equilibria and evaluating their global behavior may help us lucidly understand their inherent physical properties. 

Traditional approaches, such as Lyapunov stability theory, have provided a robust framework for addressing stability problems. However, these methods require explicit knowledge of equilibrium points and their local behavior, limiting their applicability in systems with complex dynamics. Contraction theory possesses several powerful properties for the analysis of dynamical systems \cite{lohmiller1998contraction,  jouffroy2005some, aminzare2014contraction, jouffroy2005some123}. Unlike usual Lyapunov methods to analyze the global asymptotic behavior of trajectories with respect to nominal motion, contraction ensures convergence of trajectories with respect to each other. The theory has been extensively used for observer design, synchronization, and frequency estimation \cite{aminzare2014synchronization, partialcontraction}, and \cite{sharma2008design}. It reformulates the conditions of Lyapunov stability by defining quadratic functions of differential states \cite{forni2013differential}. These functions are defined by a Riemannian contraction metric characterizing sufficient conditions for incremental exponential convergence of trajectories 
\cite{tsukamoto2021contraction, simpson2014contraction} and \cite{chaffey2018control}. The advantage of contraction-based analysis compared to the usual notion of stability is manifold. It provides powerful tools for analyzing the asymptotic behavior of time-varying nonlinear systems. The theory provides a framework for analyzing the robustness of a system to disturbances and uncertainties, an inherent property of contraction. The problem of finding an appropriate Lyapunov function is nontrivial, and it comes down to solving a partial differential equation. In contraction theory, the contraction metric for a quadratic function of the differential state can be determined by solving computationally tractable Linear Matrix Inequalities (LMIs) \cite{giesl2022review}. Another highlight is the preservation of contraction properties in the case of interconnected systems such as cascade, parallel, etc. \cite{russo2012contraction}. 

As already mentioned, contraction theory offers a complementary perspective by ensuring the convergence of neighboring trajectories to a nominal trajectory. Yet, it does not guarantee convergence to an equilibrium point, thereby necessitating additional tools for a comprehensive understanding of a system's global asymptotic behavior. For nonlinear systems with multiple isolated equilibria, the limitations of Lyapunov and classical contraction (1-contraction) theory become apparent, prompting the need for a novel analytical framework. In recent studies, there has been an attempt to solve such problems using  $2-$contraction theory \cite{RON_OFIR,9782543,WU2022110048}. The framework is a geometric generalization that enables us to study the time evolution of the volume of a $2-$dimensional parallelotope formed by two different trajectories evolving from two different initial conditions with respect to a nominal trajectory at each instant of time. This article uses the $2-$contraction theory as the backbone to analyze the dynamical systems with multiple equilibria. 

The main contributions of this paper are highlighted in the following points.
 \begin{itemize}
    \item An analytical procedure is presented to illustrate the utility of 2-contraction and Poincar$\acute{e}$ Index theory in assessing the global stability of a nonlinear system with multiple isolated equilibrium points.
    \item Analytical procedure to compute 2-contraction region for a class of planar nonlinear system has been demonstrated. Sufficient conditions are derived using 2-contraction to rule out or possible existence of closed orbit or periodic solutions in a nonlinear planar system. 
    \item A novel methodology is proposed leveraging the 2-contraction theory to approximate a common BOA for a class of planar nonlinear systems with multiple isolated stable equilibria.
    \item To highlight the efficacy, the proposed idea is implemented in a class of networked system, further extending the analysis to opinion dynamics.
 \end{itemize}

The outline of this paper is as follows: Section 2 covers key mathematical preliminaries, emphasizing contraction theory and Poincaré index theory, while also introducing the need for an alternative stability analysis approach for nonlinear systems with multiple equilibria, supported by an example. Section 3 details the proposed framework, outlining the procedure to identify the BOA for systems with multiple isolated stable equilibrium points. Section 4 illustrates several examples to validate the theorems and corollaries introduced in Section 3. Section 5 demonstrates the application of the proposed framework to networked systems, with a focus on opinion dynamics. Section 6 provides an overview of the 2D divergence theorem, explores its potential for analyzing nonlinear systems with multistability, and discusses its limitations. Finally, Section 7 concludes the paper and suggests directions for future research.

\section{Preliminaries}
\subsection{Contraction analysis}
\subsubsection{Standard contraction or 1-contraction} 
Contraction is a property that tells us about the convergence between two arbitrary trajectories of a system originating from two different sets of initial conditions. A nonlinear dynamical system is said to be contracting if initial conditions or temporary disturbances are forgotten exponentially fast. 

\begin{figure}[h!]
\begin{center}
\includegraphics[width=8cm]{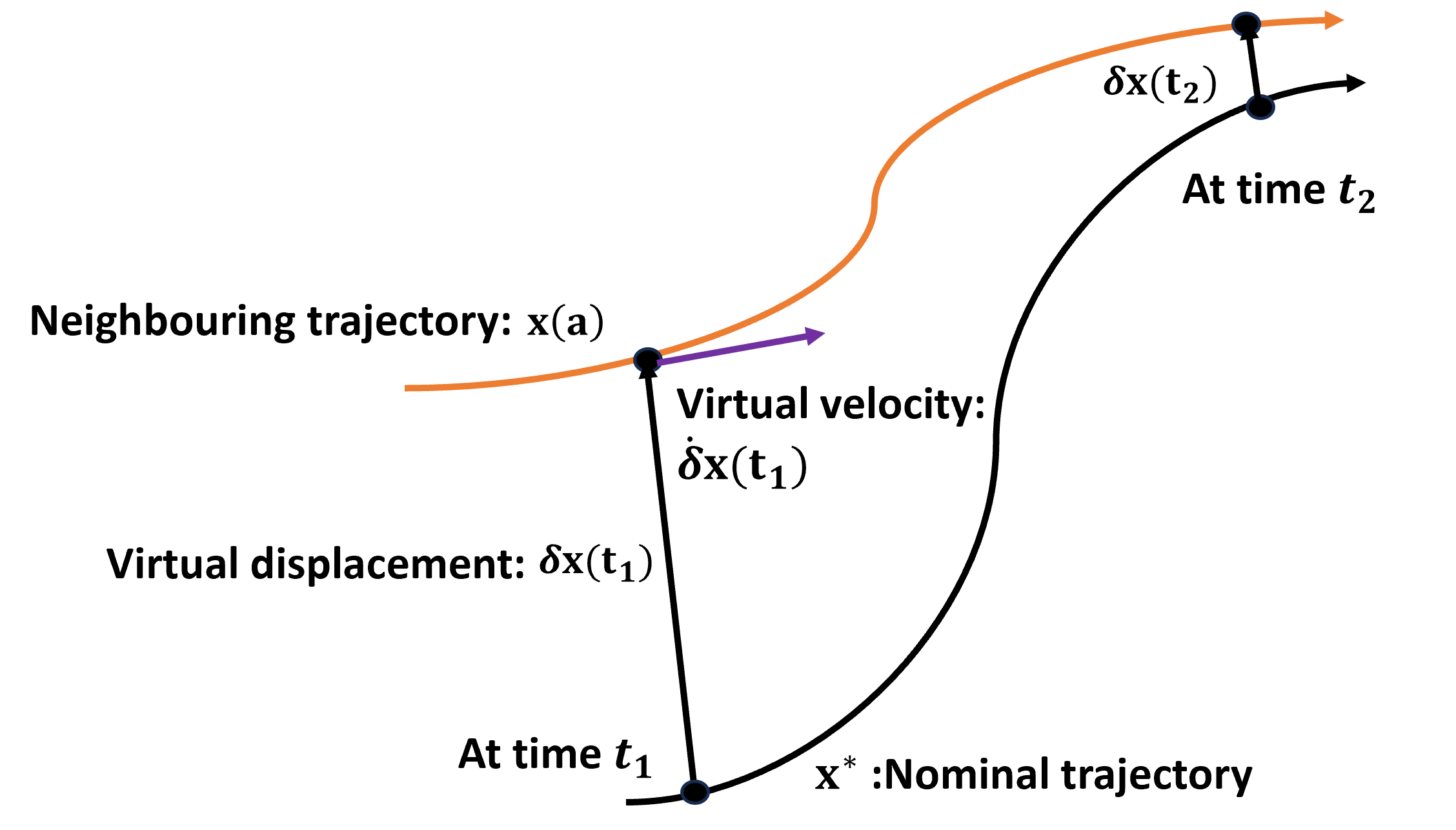}  
\caption{Evolution of virtual dynamics $\delta \mathbf{x}(t)$ with respect to time.} 
\label{Fig 2}
\end{center}
\end{figure}

\begin{figure}[h!] 
\begin{center}
\includegraphics[width=7.5cm]{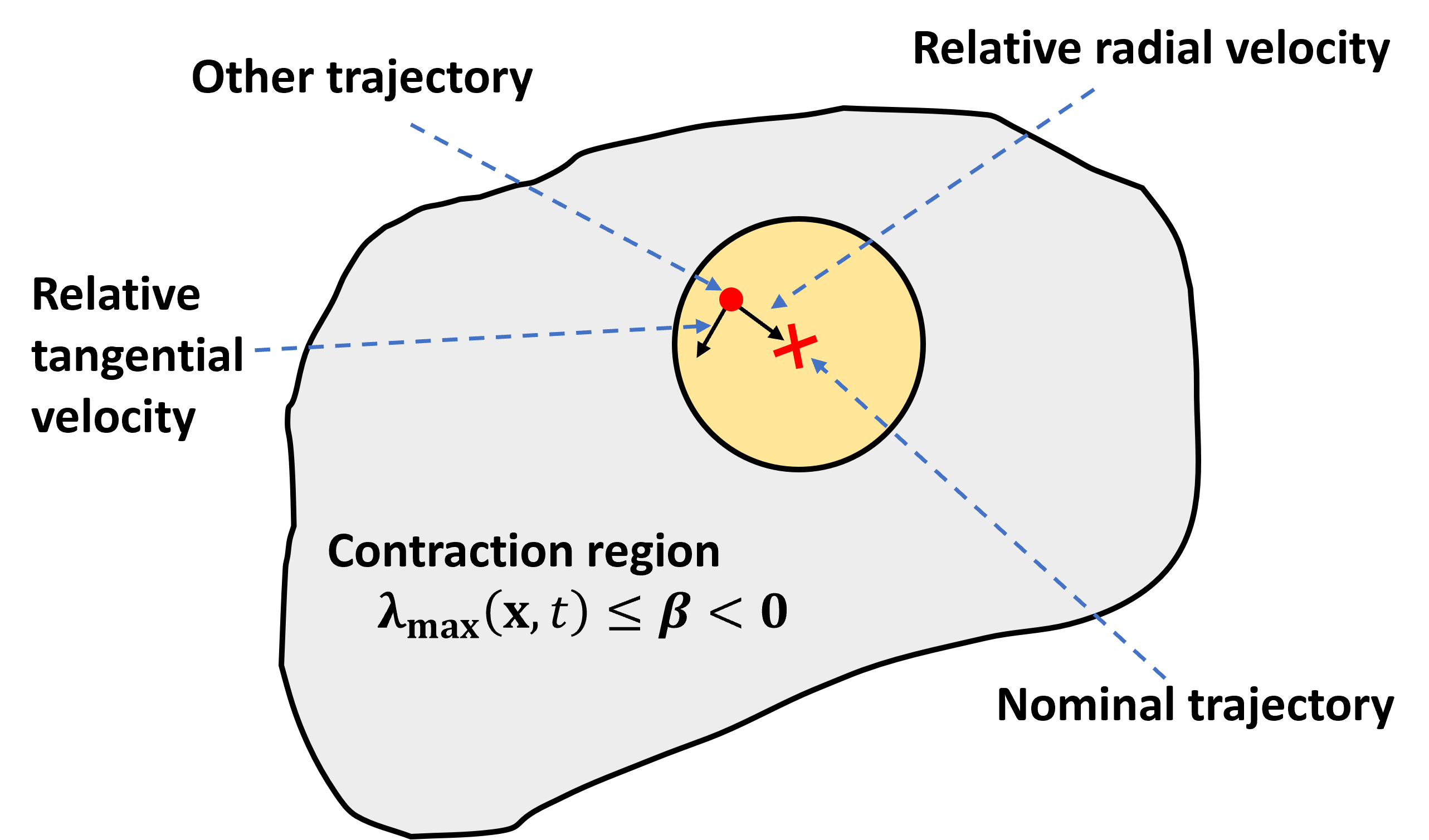}
\caption{Convergence of two trajectories in a contraction region.} 
\label{Contraction_Region}
\end{center}
\end{figure}

Consider a nonlinear dynamical system 
\begin{equation} \label{generalized nonlinear system}
\begin{split}
     \dot{\mathbf{x}}(t) & = \mathbf{f(x)}   
\end{split}
\end{equation}
where $\mathbf{f(x)} \in \Re^{n}$ is continuously differentiable function. 
We assume that the solutions evolve on a closed and convex state-space $\mathcal{E} \subset \Re^{n}$, and that for any initial condition $\mathbf{a} \in \mathcal{E}$, a unique solution $\mathbf{x(a)}$ exists and satisfies $\mathbf{x(a)} \in \mathcal{E}$ for all $t \geq 0$. Now, let $\delta \mathbf{x}(t)$ be the virtual displacement in the state $\mathbf{x}(t)$, which is the infinitesimal displacement at a fixed time. Introducing the concept of virtual dynamics, the first variations of the system can be written as:
\begin{equation} \label{Eq 9}
\begin{split}
    \delta \dot {\mathbf{x}}(t)
& = \frac{\partial \mathbf{f(x)}}{\partial \mathbf{x}}\delta \mathbf{x}(t) 
\end{split}
\end{equation}
From \eqref{Eq 9}, we drive the time derivative of the virtual distance quadratic tangent form as follows:
\begin{equation*} \label{Eq 10}
\begin{split}
\frac{d}{dt}\bigg(\delta \mathbf{x}^{T}(t)\delta \mathbf{x}(t)\bigg) & =  2 \delta \mathbf{x}^{T}(t)\frac{\partial \mathbf{f(x)}}{\partial \mathbf{x}}\delta \mathbf{x}(t) \\
      & \leq 2\lambda_{\max}(\mathbf{x},t)\delta \mathbf{x}^{T}(t)\delta \mathbf{x}(t) 
\end{split}
\end{equation*}
Assume that $\lambda_{\max}(\mathbf{x},t)$ is the maximum eigenvalue of $\frac{1}{2}\bigg(\frac{\delta \mathbf{f(x)}}{\delta \mathbf{x}} + \bigg(\frac{\delta \mathbf{f(x)}}{\delta \mathbf{x}}\bigg)^{T} \bigg)$ and is uniformly strictly negative (i.e. $\exists \beta > 0, \forall \mathbf{x}, \forall t \geq 0,$ then $\lambda_{\max}(\mathbf{x},t) \leq -\beta  < 0)$ then, any infinitesimal length $||\delta \mathbf{x}(t)||^2_2$ converges exponentially to zero. By path integration, this immediately implies that the length of any finite path converges exponentially to zero. In general, if there exists a matrix measure $\mu_{p}(.)$ (refer to Table \ref{table_matrix_measure}) with respect to some well-defined norm $p$ and $\eta_0 > 0$, such that:
 \begin{equation*}
        \begin{split}
            \mu_p (J(\mathbf{x})) \leq -\eta_0 < 0, \hspace{0.2cm} \forall t \geq 0, \forall \mathbf{x} \in \mathcal{E}
        \end{split}
    \end{equation*}
then the system \eqref{generalized nonlinear system} is called 1-contractive. This idea is pictorially represented in Fig. \ref{Fig 2}. 
\begin{definition}
 Given the system \eqref{generalized nonlinear system}, a region of the state space is called a contraction region, if the Jacobian $\frac{\partial \mathbf{f}}{\partial \mathbf{x}}$ is uniformly negative definite in that region. By $\frac{\partial \mathbf{f}}{\partial \mathbf{x}}$ uniformly negative definite we mean that
\[
\exists \beta > 0, \, \forall \mathbf{x}, \, \forall t \geq 0, \quad \frac{1}{2} \left( \frac{\partial \mathbf{f}}{\partial \mathbf{x}} + \frac{\partial \mathbf{f}}{\partial \mathbf{x}}^T \right) \leq -\beta I < 0
\]
\end{definition}
Consider now a ball of constant radius centered about a given trajectory, such that given this trajectory the ball remains within a contraction region at all times (i.e., $\forall t \geq 0$). Because any length within the ball decreases exponentially, any trajectory starting in the ball remains in the ball (since by definition the center of the ball is a particular system trajectory) and converges exponentially to the given trajectory (refer Fig. \ref{Contraction_Region}).

However, a limitation of this approach is that contraction alone does not guarantee the convergence of trajectories to an equilibrium point. It ensures that neighboring trajectories converge towards a nominal trajectory but does not necessarily lead them to an equilibrium. Therefore, to conclusively determine convergence to an equilibrium point, additional information such as the existence of equilibrium inside $\mathcal{E}$ is also required to give conclusive proof of stability. Suppose the system is contractive and has a single equilibrium point, that implies it exhibits global exponential stability (GES). Such a phenomenon is often seen in linear systems, where the local stability of the unique equilibrium point suggests global stability. But it may not be the case in nonlinear systems, because nonlinear systems often admit multiple equilibrium points. Hence, finding a global contraction region is difficult.

\begin{figure}[h!]
\begin{center}
\includegraphics[width=7cm]{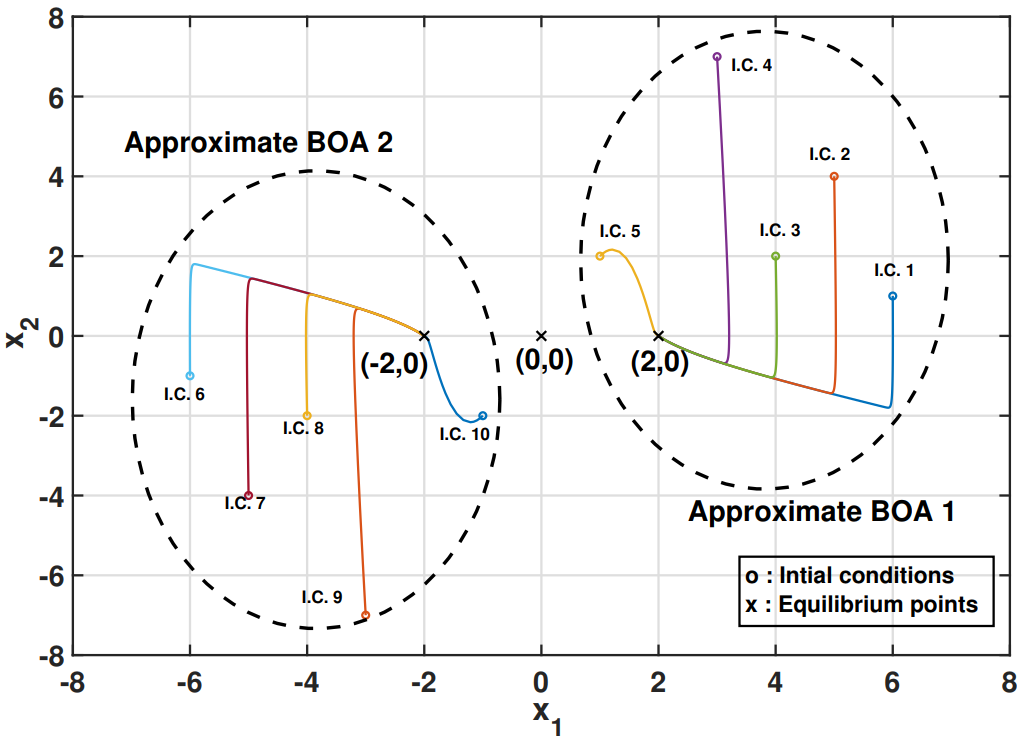}  \caption{The phase trajectories originating from 10 different initial conditions for system \eqref{2nd order nonlinear system example 1}. } 
\label{Phase Portrait Example 1}
\end{center}
\end{figure}

 Let us see a simple example of a planar nonlinear system that admits multiple equilibrium points, where 1-contraction analysis does not give global stability information. For example the following planar nonlinear system
\begin{equation} \label{2nd order nonlinear system example 1}
    \begin{split}
        \dot{x}_1 & = x_2 \\
        \dot{x}_2 & = 4x_1+ 3x_2 - x^{3}_1 - 3x^{2}_1x_2
    \end{split}
\end{equation}
has three equilibrium points: $(0,0)$, $(2,0)$, and $(-2,0)$, where $(2,0)$ and $(-2,0)$ are stable and $(0,0)$ is a saddle. Fig. \ref{Phase Portrait Example 1} shows that depending upon the choices of different initial conditions, the phase trajectory converges to an equilibrium point. Ten distinct random initial conditions are selected, as depicted in Fig. \ref{Phase Portrait Example 1}, to illustrate the behavior of phase trajectories. Trajectories originating within BOA 1 are observed to converge to the equilibrium point \((2, 0)\), while those starting in BOA 2 similarly converge to the equilibrium point \((-2, 0)\). 

\begin{definition} \label{BASIN OF ATTRACTION}
For a nonlinear system \eqref{generalized nonlinear system}, consider an attracting set $\mathcal{A} \subset \Re^n$, where $\mathcal{A}$ can be a (a) fixed point, (b) a limit cycle, (c) a more complex attractor, such as chaotic strange attractor; then the basin of attraction (BOA) can be defined as 
\begin{equation*}
    \begin{split}
        \mathcal{B}({\mathcal{A}}) & = \bigg\{\mathbf{x_0}\in \Re^n \bigg|\lim_{t\to \infty} \mathbf{x}(\mathbf{x_0},t) = \mathcal{A} \bigg\}
    \end{split}
\end{equation*}
where, $\mathbf{x}(\mathbf{x_0},t)$ is the solution (trajectory) of the system originating from initial condition $\mathbf{x_0}$.
\end{definition}

The presence of multiple stable equilibria creates separate local BOAs in the state space for each respective isolated stable equilibrium point. In other words, commenting on the steady state of the system, knowledge of the initial conditions plays a crucial role in the case of systems that exhibit multi-equilibria. 

In systems with multiple stable equilibrium points, the significance of global stability and the uniform contraction properties seen in single-equilibrium systems is no longer applicable. Instead, the dynamics are governed by local stability and localized contraction behaviors within individual basins of attraction. To extend the understanding of the global asymptotic properties of such systems, it becomes imperative to move beyond classical contraction theory. This motivates the development of \( k \)-contraction theory, particularly 2-contraction, which generalizes contraction analysis to characterize the evolution of areas (or higher-dimensional volumes) in the state space. This theory bridges the gap between local stability properties and the global asymptotic behavior of nonlinear systems. For systems with multiple equilibria identifying a common BOA ($\mathcal{A}$ contains all the stable equilibrium points) becomes highly nontrivial. We intend to propose a novel methodology leveraging the 2-contraction theory to find a common BOA for a class of nonlinear systems with multiple stable equilibria. 

\subsubsection{2-contraction}
2-contraction theory generalizes classical contraction theory by extending its focus from the contraction of trajectories (distance between solutions) to the contraction of areas in the state space. While classical contraction theory ensures that all trajectories in the state space converge exponentially to a unique trajectory or equilibrium point under a contraction metric, 2-contraction theory examines the evolution of infinitesimal two-dimensional volumes (area) formed by neighboring solutions. Specifically, in a 2-contractive system, the area spanned by any set of nearby trajectories decreases over time, eventually collapsing onto a lower-dimensional manifold. This property allows 2-contraction to capture the dynamics of systems with multiple stable equilibria, where the global contraction of distances may not hold. By quantifying how regions of the state space evolve and shrink, the 2-contraction theory provides a rigorous framework for analyzing the structure of basins of attraction, separatrices, and transitions between different attractors. 

We first define a few important terms to explore the mathematical conditions for 2-contraction of the nonlinear system \eqref{generalized nonlinear system}.

\subsubsection*{Compound matrices and matrix measures}
To define compound matrices, the following notations are used. For two integers $i,j$, with $i \leq j$, let us take $[i,j] := \{i, i+1, i+2, \ldots, j\}$. Let $Q_{k,n}$ denote the set of increasing sequences of $k$ numbers from $[1,n]$ ordered lexicographically. For example, $Q_{2,3} = \{(1,2),(1,3),(2,3) \}$. For a detailed review of set theory and lexicographic order,  refer to \cite{harzheim2005ordered}. The $I$ stands for the Identity matrix with its usual definition and appropriate size. Given any matrix, $A \in \Re^{n \times m}$ and $k \in [1, \text{min}\{n,m\}]$, the minor of order $k$ of $A$ is the determinant of some $(k \times k)$ sub-matrix of $A$. Let us define the $\frac{n!}{k!(n-k)!} \times \frac{m!}{k!(m-k)!}$ minors of order $k$ of the main matrix $A$. Each such minor is described by a set of row indices $k_i \in Q_{k,n}$ and column indices $k_j \in Q_{k,m}$. This minor is depicted by $A(k_i|k_j)$. For detailed explanations of compound matrices, the reader is suggested to refer \cite{deutsch1975matrix, martin1974logarithmic}. 
\subsubsection*{Computation of compound matrix}
\begin{equation*}
    \begin{split}
\text{Let} \hspace{0.2cm} A & = 
\left[ \begin{array}{rr}
1 & 6\\
-7 & 0 \\
5 & 10
\end{array}\right]_{3 \times 2} \text{so,} \hspace{0.2cm} A(\{1,3\}|\{1,2\})=\det\left[ \begin{array}{rr}
1 & 6\\
5& 10
\end{array}\right] 
    \end{split}
\end{equation*}
For $n=m=3$ and $k=2$, the multiplicative compound matrix of $A$:
\small
$$
A^{(2)} = \left[ \begin{array}{rrr}
A(\{1,2\}|\{1,2\}) & A(\{1,2\}|\{1,3\}) & A(\{1,2\}|\{2,3\})\\
A(\{1,3\}|\{1,2\}) & A(\{1,3\}|\{1,3\}) & A(\{1,3\}|\{2,3\}) \\
A(\{2,3\}|\{1,2\}) & A(\{2,3\}|\{1,3\}) & A(\{2,3\}|\{2,3\})
\end{array}\right]
$$ 
\normalsize
By definition, we can find some important relation, such that: $A^{(1)}=A$, and if $A \in \Re^{n \times n}$ then $A^{(n)} = \det (A)$.

It is important to note that any entry in $A^{(k)}$ is a polynomial in the entries of $A$. The $k^{th}$ additive compound matrix of $A \in \Re^{n \times n}$ is defined by $A^{[k]} := \frac{d}{d \epsilon}[I+ \epsilon A]^{(k)}\Big|_{\epsilon = 0}$. This implies
$(I + \epsilon A)^{(k)} = I + \epsilon A^{[k]} + HOT(\epsilon)$; where $\epsilon A^{[k]}$ is the first-order term in the Taylor series expansion of $(I + \epsilon A)^{(k)}$.

Let us now consider the case where $n=2$ and $k=2$. Then the procedure to calculate $A^{[2]}$ additive compound matrix of a $(2 \times 2)$ matrix $A$ is depicted as shown below:
\small
\begin{equation*}
A = 
\left[ \begin{array}{rr}
a_{11} & a_{12}\\
a_{21} & a_{22}
\end{array}\right] \Rightarrow (I+\epsilon A)^{(2)}  = 
\left[ \begin{array}{rr}
1+ \epsilon a_{11} & \epsilon a_{12}\\
\epsilon a_{21} & 1+ \epsilon a_{22}
\end{array}\right]^{(2)}    
\end{equation*}
We find,
\begin{equation*}
    \begin{split}
 A^{[2]} & =\frac{d}{d \epsilon}\bigg[(I+\epsilon A)^{(2)}\bigg]_{\epsilon=0}=a_{11} + a_{22} = \text{Trace}(A)     
    \end{split}
\end{equation*}
\normalsize
$A^{[k]}$ can be explicitly described in terms of the entries $a_{ij}$ of the original matrix $A$. 
\subsubsection*{Matrix measures}
Let us consider a vector norm $||.|| : \Re^{n} \rightarrow \Re_{+}$. The induced matrix norm $||.|| : \Re^{n \times n} \rightarrow \Re_{+}$ is $||A|| := \max_{||\mathbf{x}||=1}||A.\mathbf{x}||$ and the induced matrix measure $\mu(.): \Re ^{n \times n} \rightarrow \Re$ is $\mu(A) := \lim_{\epsilon \to 0^+}\frac{||I+\epsilon A|| - 1}{\epsilon}$.

Based on the corresponding norm considered, the matrix measure definition changes which can be referred to in the following table \cite{vidyasagar2002nonlinear}.
\begin{table}[h!]
\centering
\small 
\renewcommand{\arraystretch}{1.5} 
\begin{tabular}{|c|c|}  
\hline
Vector Norms $||(.)||$ & Matrix Measure $\mu(A)$\\ 
\hline 
$ ||\mathbf{x}||_{1}=\sum_{i=1}^{n}{|x_{i}|}$ & $\mu_{1}(A)=\max_{j}{(a_{jj}}+\sum_{i\neq j}{|a_{ij}|})$\\
\hline
$ ||\mathbf{x}||_{2}=(\sum_{i=1}^{n}{|x_{i}|^{2}})^\frac{1}{2}$ & $\mu_{2}(A)=\lambda_{max}(\frac{A+A^{T}}{2})$\\
\hline
$ ||\mathbf{x}||_{\infty}=\max_{1\leq i \leq n}{|x_{i}|}$ & $\mu_{\infty}(A)=\max_{i}{(a_{ii}}+\sum_{i\neq j}{|a_{ij}|})$\\
\hline
\end{tabular}
\caption{Matrix Measures Corresponding to Vector Norms.}
\label{table_matrix_measure}
\end{table}

Again, the matrix measures for any additive compound matrix A$^{[k]}$ can be formulated as shown \cite{WU2022110048}:
\small
\begin{equation*} \label{Eq 4}
    \begin{split}
       \mu_1(A^{[k]}) & = \text{max}_{i}\bigg(\sum_{p=1}^{k}a_{i_p,i_p}+\sum_{j \notin i}\bigg(|a_{j,i_1}|+\ldots+|a_{j,i_k}|\bigg)\bigg)\\
       \mu_2(A^{[k]}) & = \sum_{i=1}^{k}\lambda_i\bigg(\frac{A+A^{T}}{2}\bigg) \\
       \mu_\infty(A^{[k]}) & = \text{max}_{i}\bigg(\sum_{p=1}^{k}a_{i_p,i_p}+\sum_{j \notin i}\bigg(|a_{i_1,j}|+\ldots+|a_{i_k,j}|\bigg)\bigg)\\
    \end{split}
\end{equation*}
\normalsize
the maximum is taken over all $k-$tuples, where $i=\{i_1,i_2,\ldots,i_k\}$.

Matrix measures, or logarithmic norms, play a crucial role in dynamical system analysis as they provide insights into system stability, robustness, and convergence behavior. For linear systems, matrix measures characterize the exponential growth or decay of solutions, directly indicating stability (e.g., $\mu(A) < 0$ implies asymptotic stability). In nonlinear systems, they are used in conjunction with the Jacobian to analyze local stability near equilibrium points. Matrix measures also facilitate contraction analysis, where negative values indicate that trajectories converge over time. They are essential for bounding solutions.

\subsubsection*{Mathematical formulation of 2-contraction}
\begin{figure}[h!]
\begin{center}
\includegraphics[width=8.5cm]{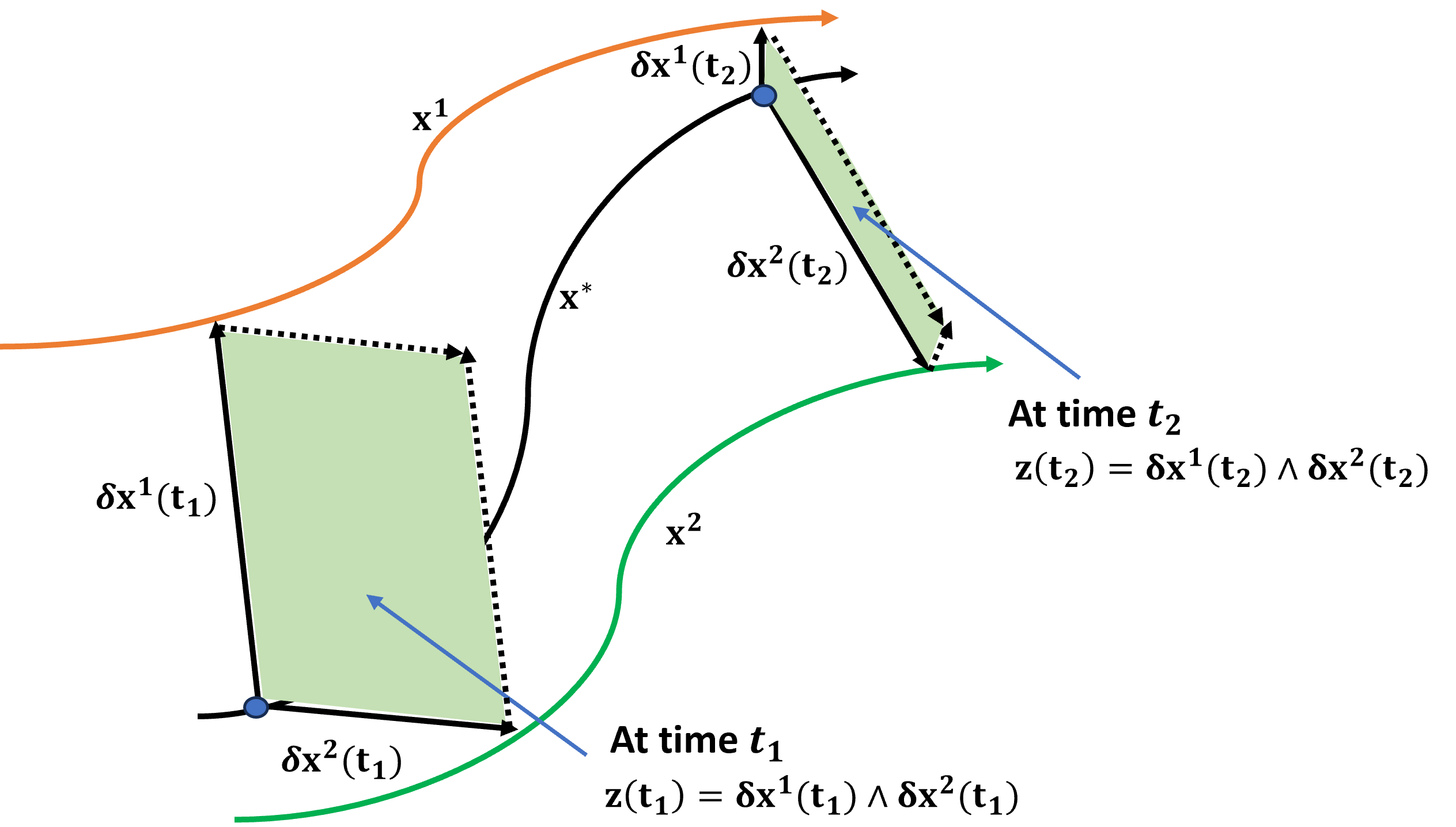}  
\caption{Evolution of area $z(t)$ with respect to time. }  
\label{Virtual dynamics of the infinitesimal area. }
\end{center}
\end{figure}

Consider $\mathbf{x^1}$ and $\mathbf{x^2}$ are two neighboring trajectories originating from two different initial conditions with respect to the nominal trajectory $\mathbf{x^*}$. For the $n-$dimensional system \eqref{generalized nonlinear system}, define $n-$infinitesimal vectors $\delta \mathbf{x^1}=\mathbf{y^1}$ and $\delta \mathbf{x^2}=\mathbf{y^2}$ in the tangent space with respect to its nominal trajectory $\mathbf{x^*}$. The infinitesimal area vector spanned by $\mathbf{y^1}$ and $\mathbf{y^2}$ is defined as 
\begin{equation*}
    \begin{split}
        \mathbf{z} & =\mathbf{y^1} \wedge \mathbf{y^2} = \bigg[ \mathbf{y^1} \hspace{0.2cm} \mathbf{y^2}\bigg]^{(2)}
    \end{split}
\end{equation*}
Defining variational dynamics of the virtual area is (refer to Fig. \ref{Virtual dynamics of the infinitesimal area. }) 
\begin{equation}
    \begin{split}
        \dot{\mathbf{z}} & = J^{[2]} \mathbf{z}
    \end{split}
\end{equation}
The magnitude of the virtual area vector is $S = \mathbf{z}^T\mathbf{z} = \parallel \mathbf{z} \parallel^2_2$. So, the evolution of $S$ with respect to time can be defined as follows:
\begin{equation*}
    \begin{split}
    \dot{S} & = 2S\bigg(\frac{(J^{[2]})^T+J^{[2]}}{2}\bigg)S
    \end{split}
\end{equation*}
\begin{equation} \label{2contractioncodntion}
    \begin{split}
    \Rightarrow \dot{S} & \leq 2\mu_2(J^{[2]}) S
    \end{split}
\end{equation}
The system \eqref{generalized nonlinear system} is said to be 2-contractive; i.e. the infinitesimal area formed by two trajectories originating from two different initial conditions at each time $t$ with respect to a nominal trajectory $\mathbf{x^*}$ will shrink to zero as $t \to \infty$ if in the condition \eqref{2contractioncodntion} $\mu_2(J^{[2]}) < 0$.

A few key results that are tightly linked with 2-contraction analysis are highlighted as follows:
\begin{lemma} \cite{WU2022110048} \label{2contraction theorem}
If there exists a scalar $\eta > 0$ and a vector norm $\parallel . \parallel$, with induced matrix measure $\mu: \Re^{n \times n} \to \Re$, such that 
    \begin{equation} \label{2-contractive condition}
        \begin{split}
            \mu (J^{[2]}(\mathbf{x})) \leq -\eta < 0, \hspace{0.2cm} \forall t \geq 0, \forall \mathbf{x} \in \mathcal{E}
        \end{split}
    \end{equation}
\end{lemma}
then \eqref{generalized nonlinear system} is 2-contractive. 

\begin{lemma} \cite{li1995ra} \label{Theorem 2.5}
Consider the nonlinear system \eqref{generalized nonlinear system}, where $f: \Re^n \to \Re$ is $C^1$. Assume that its trajectories evolve in a convex and compact set $\mathcal{E}$ and that $\mu(J^{[2]})<0$ for all $\mathbf{x} \in \mathcal{E}$. Then every solution emanating from $\mathcal{E}$ converges to the set of equilibria.  
\end{lemma}

\begin{lemma}
 \label{no periodic solution} \cite{WU2022110048}
    Suppose that one of the inequalities 
    \begin{equation*}
        \begin{split}
            \mu(J^{[2]}(\mathbf{x}))<0 \hspace{0.2cm} \text{or} \hspace{0.2cm} \mu(J^{[2]}(\mathbf{x}))>0 \hspace{0.2cm}
        \end{split}
    \end{equation*}
    holds for all $\mathbf{x} \in \Re^n$. Then the system \eqref{generalized nonlinear system} has no non-constant periodic solutions.
\end{lemma}

\subsection{Poincar$\acute{\textbf{e}}$ Index theory \cite{meiss2007differential,strogatz2001nonlinear}} \label{Index Theory}
 Local analysis, such as linearizing a system around an equilibrium point and examining the eigenvalues of the Jacobian, provides a microscopic view of the trajectories near the equilibrium. However, it cannot predict the behavior of trajectories once they exit the local neighborhood \cite{strogatz2001nonlinear}. Index theory analyzes the behavior of a vector field over a simple closed curve $C$ (i.e. does not intersect itself), providing global insights into the system's properties, such as the total number and nature of equilibrium points, that are enclosed by $C$, without requiring local analysis. We intend to utilize this idea to comment on the behavior of vector fields or the nature of equilibrium points of a nonlinear planar system that exhibits multiple equilibrium points, without explicitly doing local analysis.

\begin{figure}[h!]
\begin{center}
\includegraphics[width=5cm]{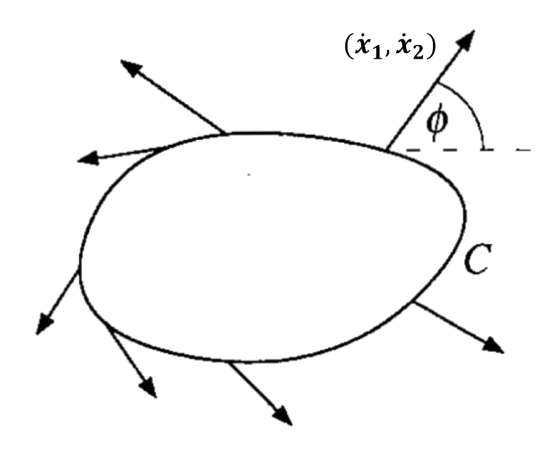}  
\caption{Definition of the Poincar$\acute{e}$ index. } 
\label{index_theory_figure}
\end{center}
\end{figure}

Define a planar system \eqref{generalized nonlinear system} for $n = 2$ with $\mathbf{f}(\mathbf{x})$ to be $C^1$. Consider $C$ (refer to Fig. \ref{index_theory_figure}) that does not pass through any equilibrium points of the system. $C$ is not necessarily to be a trajectory, it is simply a loop that has been put in the phase plane to probe the behavior of the vector field. The index of a closed curved $C$ is an integer that measures the winding of the vector field on $C$. Then at each point $\mathbf{x}$ on $C$, the vector field $\dot{\mathbf{x}}=[f_1 \hspace{0.2cm} f_2]^T$ makes a well-define angle $\Phi = \text{tan}^{-1}\bigg(\frac{f_2}{f_1}\bigg)$ with the positive $x_1-$axis (refer to Fig. \ref{index_theory_figure}).

The index $I_C$ is given by:
\begin{equation} \label{POINCARE INDEX FORMULA}
    \begin{split}
       I_C(\mathbf{f}) & = \frac{\Delta\Phi}{2\pi} 
    \end{split}
\end{equation}
where, $\Delta \Phi$ is the net change in the direction of $\mathbf{f}$ upon traversal of the loop. When $\mathbf{f}$ is $C^1$, $\Delta \Phi$ can be obtained by integrating along the curve:
\begin{equation} \label{eq 8}
    \begin{split}
        I_C(\mathbf{f}) & = \frac{1}{2 \pi}\oint_C d\Phi = \frac{1}{2\pi} \oint_C \frac{f_1 df_2 - f_2df_1}{f^2_1 + f^2_2}
    \end{split}
\end{equation}
This can be computed easily if the $C$ is written in parametric form. Few important observations of $I_C(\mathbf{f})$ in \eqref{eq 8} are summarized \cite{strogatz2001nonlinear}:
\begin{itemize}
    \item If $C$ encloses a stable or an unstable equilibrium point then $I_C(\mathbf{f}) = +1$. 
    \item If $C$ encloses a closed orbit or a non-constant periodic solution then $I_C(\mathbf{f}) = +1$.
    \item If $C$ encloses a saddle equilibrium point then $I_C(\mathbf{f}) = -1$. 
    \item If $C$ does not enclose any equilibrium points, $I_C(\mathbf{f}) = 0$.
\end{itemize}
The summary of the index theory is stated in the form of two lemmas as follows:
\begin{lemma}\label{Index Thoery for Mult. Fixed Points} \cite{strogatz2001nonlinear}
If a closed curve $C$ surrounds $N$ isolated equilibrium points $\mathbf{x^*_1,x^*_2,\ldots,x^*_N}$, then the total index 
\begin{equation} \label{Total index around C}
    \begin{split}
        I_C(\mathbf{f}) & = \sum_{i=1}^{N} I_i(\mathbf{f})
    \end{split}
\end{equation}
where, $I_i$ is the index of $\mathbf{x^*_i}$, for $i=1,2,\ldots,N$.
\end{lemma}



\section{Finding a common BOA for a nonlinear system with multiple equilibria}
Let us consider a class of nonlinear planar system 
\begin{equation} \label{general nonlinear planar system}
    \begin{split}
\dot{\mathbf{x}} & = A\mathbf{x}+\mathbf{f(x)}
     \end{split}
\end{equation}
with; 
\begin{equation*}
A = 
\left[ \begin{array}{rr}
0 & 1\\
b_1 & b_2
\end{array}\right], \mathbf{f(x)}=
\left[ \begin{array}{r}
0\\
-b_3x^{2s+1}_1 + b_4x^{2m}_1x^{2q+1}_2
\end{array}\right]
\end{equation*}
where $\mathbf{x} = [x_1 \hspace{0.2cm} x_2] \in \Re^2$ is the state vector, $\mathbf{f}: \Re^2 \to \Re^2$ is assumed to be of class $C^1$, $\mathbf{x}(t)$ is the state time-evolution with initial condition $\mathbf{x(0)} = \mathbf{x_0}$. Let $\mathbf{f(x)}$ satisfies $\mathbf{f}(x_1,0) = -\mathbf{f}(-x_1,0), \mathbf{f(0)} = \mathbf{0}$. Also, $\{m,q\} \in \mathbf{W}, s \in \mathbf{N}$ and $b_i \in \Re, i=1,2,3,4$;$b_j >0, j =\{1,3\}$. System \eqref{general nonlinear planar system} has three equilibrium points, $\mathbf{x^*_1}: (0,0)$, $ \mathbf{x^*_2}: \bigg(\frac{b_1}{b_3}^{\frac{1}{2s}},0\bigg)$, $\mathbf{x^*_3}: \bigg(-\frac{b_1}{b_3}^{\frac{1}{2s}},0\bigg)$. The position of the equilibrium points is independent of the parameters $\{m,q,b_2,b_4\}$. 

The class of system \eqref{general nonlinear planar system} is commonly found and studied in various fields of physics, engineering, biology, and applied mathematics \cite{strogatz2001nonlinear,rao1995mechanical}. 
\subsection{Computation of 2-contraction region}
In this section, we formally define the 2-contraction region for system \eqref{general nonlinear planar system} in the form of Proposition \ref{Theorem to find Omega}.

\begin{proposition} \label{Theorem to find Omega}
Given the system \eqref{general nonlinear planar system} there exists a 2-contractive region $\Omega \subseteq \Re^2$; defined as
   \small
    \begin{equation} \label{2 contraction region}
        \begin{split}
      \Omega & = \bigg\{(x_1,x_2) \in \Re^2 \bigg| b_2+b_4(2q+1)x^{2m}_1x^{2q}_2 < 0\bigg\} 
        \end{split}
    \end{equation}
\end{proposition}
\begin{proof}
\normalsize
For a general nonlinear system \eqref{generalized nonlinear system} to be 2-contractive condition (\ref{2-contractive condition}) should be satisfied. For the class of system \eqref{general nonlinear planar system}, Jacobian  $J = \bigg( A + \frac{\partial \mathbf{f(x)}}{\partial \mathbf{x}} \bigg)$, hence $\mu_2(J^{[2]}) = \text{Trace}(J) = \text{Trace}\bigg( A + \frac{\partial \mathbf{f(x)}}{\partial \mathbf{x}} \bigg) \leq -\eta < 0, \eta > 0, \forall \mathbf{x} \in \Re^2$. For the system \eqref{general nonlinear planar system} to be 2-contractive in the region $\Omega$ condition \eqref{2 contraction region} should be satisfied. $\blacksquare$
\end{proof}

\begin{remark}
   Based on condition (\ref{2 contraction region}) we can define two other regions $\Omega_0$ and $\Omega_1$ in the state space. These regions are defined as
 \small
 \begin{equation} \label{Omega_1 and Omega_0}
        \begin{split}
       \Omega_1 & = \bigg\{(x_1,x_2) \in \Re^2 \bigg|b_2+b_4(2q+1)x^{2m}_1x^{2q}_2 > 0\bigg\}  \\
       \Omega_0 & = \bigg\{(x_1,x_2) \in \Re^2 \bigg| b_2+b_4(2q+1)x^{2m}_1x^{2q}_2 = 0\bigg\}.  \\
        \end{split}
    \end{equation} 
\end{remark}
\normalsize
In \(\Omega_1\), \(J^{[2]} > 0\) implies \(\mu_2(J^{[2]}) > 0\). As a result, based on Lemma \ref{no periodic solution}, \(\Omega_1\) cannot contain any non-constant periodic solutions. Suppose $\lambda_i(\mathbf{x}), i=1,2$ are two state-dependent eigenvalues of the Jacobian matrix of the original system $\eqref{general nonlinear planar system}$ and in $\Omega_1$, $J^{[2]}$ = Trace$(J(\mathbf{x})) = \lambda_1(\mathbf{x})+\lambda_2(\mathbf{x}) > 0, \forall t \geq 0$, so no stable equilibrium points lie in $\Omega_1$. Hence, the possible natures of equilibrium point that exist in $\Omega_1$ are (i) saddle and/or  (ii) Unstable. On the other hand, $\Omega_0$ is the boundary between $\Omega_1$ and $\Omega$ where $\mu_2(J^{[2]}) = 0$. If any equilibrium points lie on $\Omega_0$, non-constant periodic trajectories can emerge in their vicinity. So it is assumed that no equilibrium points lie on $\Omega_0$ (in case there exists a $\Omega_0$). The overall behavior of the system dynamics can be accessed by combining the information obtained from $\Omega_1, \Omega_0$ and $\Omega$.

\subsection{Nature of equilibrium points}
The nature of each equilibrium point of \eqref{general nonlinear planar system} is assumed to be unknown, and we avoid local analysis, such as linearizing the system around individual equilibrium points. In this section the application of Lemma \ref{Index Thoery for Mult. Fixed Points} has been shown to characterize the equilibrium points. System \eqref{general nonlinear planar system} has three equilibrium points $\mathbf{x^*_1},\mathbf{x^*_2}$ and $\mathbf{x^*_3}$, i.e. $N=3$. Now, consider a counterclockwise oriented simple closed circular curve $C = \bigg\{(x_1,x_2) \in \Re^2 \bigg|x^2_1+x^2_1 = r_0 \bigg\}$. The positive scalar $r_0$ is chosen such that it encloses all the equilibrium points. Using Lemma \ref{Index Thoery for Mult. Fixed Points} assume that, $I_C(\mathbf{f}) = I_1(\mathbf{f})+I_2(\mathbf{f})+I_3(\mathbf{f})= +1$. Then the only possible combination is $I_C = -1 + 1 +1$. It indicates that there must be one saddle (i.e. index -1) inside $C$. Let us consider $\mathbf{x^*_1}$ be the saddle equilibrium point, then $\mathbf{x^*_1} \in \Omega_1$ as defined in \eqref{Omega_1 and Omega_0}. As already assumed, no equilibrium points lie on $\Omega_0$. So, $\mathbf{x^*_2}$ and $\mathbf{x^*_3}$ must lie in the $\Omega$ region. Based on Lemma \ref{no periodic solution}, $\Omega$ does not include any non-constant periodic trajectories. So, these equilibrium points may be stable or unstable. As per the definition of condition given in \eqref{2 contraction region}, $\Omega$ does not contain any unstable equilibrium points. Hence, $\{\mathbf{x^*_2},\mathbf{x^*_3}\}\in \Omega$ have to be stable. Suppose $I_C(\mathbf{f})= +1$ and there does not exist $\Omega_1$ and $\Omega_0$ regions for system \eqref{general nonlinear planar system} then all the equilibrium points $\mathbf{x^*_i}, i=1,2,3$ lie in $\Omega$. On the other hand, suppose $I_C(\mathbf{f})=+1$ and there does not exist $\Omega$ region, all the equilibrium points $\mathbf{x^*_i}, i=1,2,3$ must lie in $\Omega_1$. These aspects are also discussed in the examples discussed subsequently.

\subsection{Definition of common BOA for a planar system}
The following definition \ref{Common BOA Definition} is introduced to formally define the common BOA for the generalized nonlinear planar system \eqref{generalized nonlinear system} when $n = 2$.

\begin{definition} \label{Common BOA Definition} 
An open set $\mathcal{D}_0 \subseteq \Re^2$ is a common BOA for the planar system $\dot{\mathbf{x}}=\mathbf{f}(\mathbf{x})$ such that $\mathbf{x(0)} \in \mathcal{D}_0$ then $\mathbf{x}(t) \to \mathbf{x^*_i}; i=\{1,2,\ldots,N\}$ as $t \to \infty$, where $\mathbf{x^*_i}$ is any one of the isolated equilibrium points from $\{\mathbf{x^*_1},\mathbf{x^*_2},\ldots,\mathbf{x^*_N}\}$ in $\mathcal{D}_0$.
\end{definition}

\subsection{General steps to compute common BOA }\label{section 3.3}
In this section, we provide the required steps for computing the common BOA for a general nonlinear planar system. Later we adopt these steps for characterizing a candidate $\mathcal{D}_0$ for different classes of systems.

\begin{itemize}

\item \textbf{Step 1:} Identify the region $\Omega$ in $\Re^2$ state space that satisfies the 2-contraction condition (\ref{2-contractive condition}) (refer to Lemma \ref{2contraction theorem}). If $\Omega$ exists, then proceed to Step 2. 

\item \textbf{Step 2:} Select $C$ such that it encloses all the isolated equilibrium points admitted by the planar system. Identify the nature of isolated equilibrium points using Lemma \ref{Index Thoery for Mult. Fixed Points}. If $I_C(\mathbf{f}) = +1$, then proceed to step 3.

\item \textbf{Step 3:} To satisfy Lemma \ref{Theorem 2.5}, boundedness of the system trajectories is to be ensured. Define a candidate energy function $E(\mathbf{x})$ for the system. Then find out the region $\mathcal{U}$, where $\frac{d E(\mathbf{x})}{dt} \leq 0, \forall t \geq 0$.

\item \textbf{Step 4:} Finally, the common BOA is defined as $\mathcal{D}_0 = \Omega \bigcap \mathcal{U}$.
\end{itemize}

We provide Theorem \ref{Define D0 region} to compute $\mathcal{D}_0$ for system \eqref{general nonlinear planar system} adopting the algorithm given in this section.

\begin{figure}[h!]
\begin{center}
\includegraphics[width=7cm]{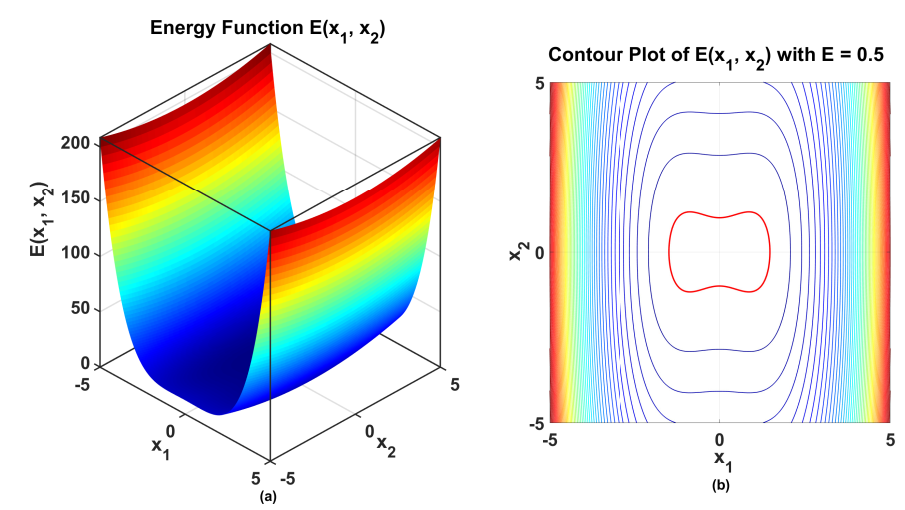} 
\caption{Energy function $E(\mathbf{x})$ \eqref{energy_E(X)} is plotted with level set curves for the system \eqref{general nonlinear planar system} with $b_1 = 1,b_2 =-1,b_3=1,b_4 =0, s =1$. (a) Energy function $E(\mathbf{x})$, (b)The level set curve (red colored curve) for $r = 0.5$. } 
\label{enery_level_set}
\end{center}
\end{figure}

\begin{theorem}\label{Define D0 region}
 Suppose, for system \eqref{general nonlinear planar system} there exists a region $\Omega$ and also a simple closed curve $C$ such that $I_C(\mathbf{f}) = +1$. Then there exists a candidate energy function $E(\mathbf{x})$ given in \eqref{energy_E(X)} 
\begin{equation} \label{energy_E(X)}
    \begin{split}
        E(\mathbf{x}) & = \frac{1}{2}x_2^2-\frac{b_1}{2}x_1^2+\bigg(\frac{b_3}{2s+2}\bigg)x^{2s+2}_1
    \end{split}
\end{equation}
with level set $\partial \mathcal{U} = \bigg\{(x_1,x_2) \in \Re^2\bigg| E(\mathbf{x}) = r\bigg\}$ and
\begin{equation} \label{U region}
    \begin{split}
        \mathcal{U} & = \bigg \{(x_1,x_2)\in \Re^2\bigg| E(\mathbf{x}) < r\bigg\}
    \end{split}
\end{equation}
such that $\frac{d E(\mathbf{x})}{dt} \leq 0$ in $\mathcal{U}$. Then common basin of attraction $\mathcal{D}_0$ is $\mathcal{D}_0 = \Omega \bigcap \mathcal{U}$, where $\Omega$ is the 2-contraction region defined in \eqref{2 contraction region}. The scalar $r$ can be selected to adjust the overall shape of $\mathcal{U}$.
\end{theorem} 
\proof
To prove that $\mathcal{U}$ is bounded, let us define a candidate energy function $E(\mathbf{x})$ which is continuously differentiable in $\Re^2$. The choice of $E(\mathbf{x})$ is inspired by \emph{double well potential function theory} \cite{strogatz2001nonlinear,zhu2013stochastic}. For the class of system \eqref{general nonlinear planar system} we define $E(\mathbf{x})$ as given in \eqref{energy_E(X)}.
We evaluate
\begin{equation} \label{compactness of U}
    \begin{split}
        \frac{d E}{dt} & = b_2x^2_2 + b_4 x^{2m}_1 x^{2q+2}_2 < 0
    \end{split}
\end{equation}
The region $\mathcal{U}$ given in \eqref{U region} is said to be bounded if at each point in $\mathcal{U}$ condition (\ref{compactness of U}) is satisfied. This implies that the level set $\partial\mathcal{U} = \bigg\{(x_1,x_2) \in \Re^2 \bigg|E(\mathbf{x}) = r\bigg\}$ (refer to Fig. \ref{enery_level_set})  is a closed and bounded region, that means $\mathbf{x(0)} \in \mathcal{U} \to \mathbf{x}(t) \in \mathcal{U}, \forall t \geq 0$. 

According to Lemma \ref{no periodic solution}, $\Omega$ region does not contain any closed orbit and non-constant periodic solutions. Also, $\mathcal{U}$ is proved to be a bounded region. Now the common BOA for system \eqref{general nonlinear planar system} is $\mathcal{D}_0 = \Omega \bigcap \mathcal{U}$. So, according to Lemma \ref{Theorem 2.5} if $\mathbf{x}(\mathbf{0}) \in \mathcal{D}_0$ then $\mathbf{x}(t)$ converge to one of the equilibrium points as $t \to \infty$. $\blacksquare$

\vspace{-1cm}
\section{Application to benchmark examples}
In this section, the application of the Theorem \ref{Define D0 region} is shown with the help of three benchmark examples in order to highlight different cases.

\subsection{Example 1: A system with $\Omega = \Re^2$}
The parameters for system \eqref{general nonlinear planar system} are chosen as follows: $b_1 = 1, b_2 = -1 < 0, b_3 = 1, b_4 = 0$ and $s = 1$.
\begin{equation} \label{First example}
\dot{\mathbf{x}} = 
\left[ \begin{array}{rr}
0 & 1\\
1 & -1
\end{array}\right]\mathbf{x} + 
\left[ \begin{array}{r}
0\\
-x^3_1
\end{array}\right]
\end{equation}
The equilibrium points of system \eqref{First example} are $\mathbf{x^*_1}=(0,0)$, $\mathbf{x^*_2} = (1,0)$ and $\mathbf{x^*_3} = (-1,0)$. The 2-contraction region is computed as
\begin{equation} \label{ALL THE REGIONS}
    \begin{split}
       \Omega & = \bigg\{(x_1,x_2)|x_1 \in \Re, x_2 \in \Re\bigg\} 
    \end{split}
\end{equation}
which suggests that entire $\Re^2$ space is globally 2-contractive. So $\Omega_0$ and $\Omega_1$ do not exist. So, all the equilibrium points $\mathbf{x^*_1},\mathbf{x^*_2}$ and $\mathbf{x^*_3}$ are contained in $\Omega$.   

Consider a counterclockwise oriented simple closed circular curve $C = \bigg\{(x_1,x_2) \in \Re^2 \bigg|x^2_1+x^2_1 =16 \bigg\}$ that encloses all the equilibrium points. Applying the index theory, we get $I_C = +1$ (refer to Appendix A for detailed calculation). Based on the analysis given in Section 3.2, we can conclude that one equilibrium point is saddle and the rest of the two equilibrium points are stable in nature. 

According to Theorem \ref{Define D0 region}, $\mathcal{U}$ is given as 
\begin{equation*}
    \begin{split}
 \mathcal{U} & = \bigg \{(x_1,x_2)\in \Re^2\bigg|\frac{1}{2}x_2^2-\frac{1}{2}x_1^2 +\frac{1}{4}x^4_1 < r\bigg\}    
    \end{split}
\end{equation*}
where, $E(\mathbf{x})= \frac{1}{2}x^2_2 - \frac{1}{2}x^2_1 + \frac{1}{4}x^4_1$ and $\frac{dE(\mathbf{x})}{dt} \leq 0, \forall t \geq 0$ in $\mathcal{U}$.

\begin{figure}[h!]
\begin{center}
\includegraphics[width=7cm]{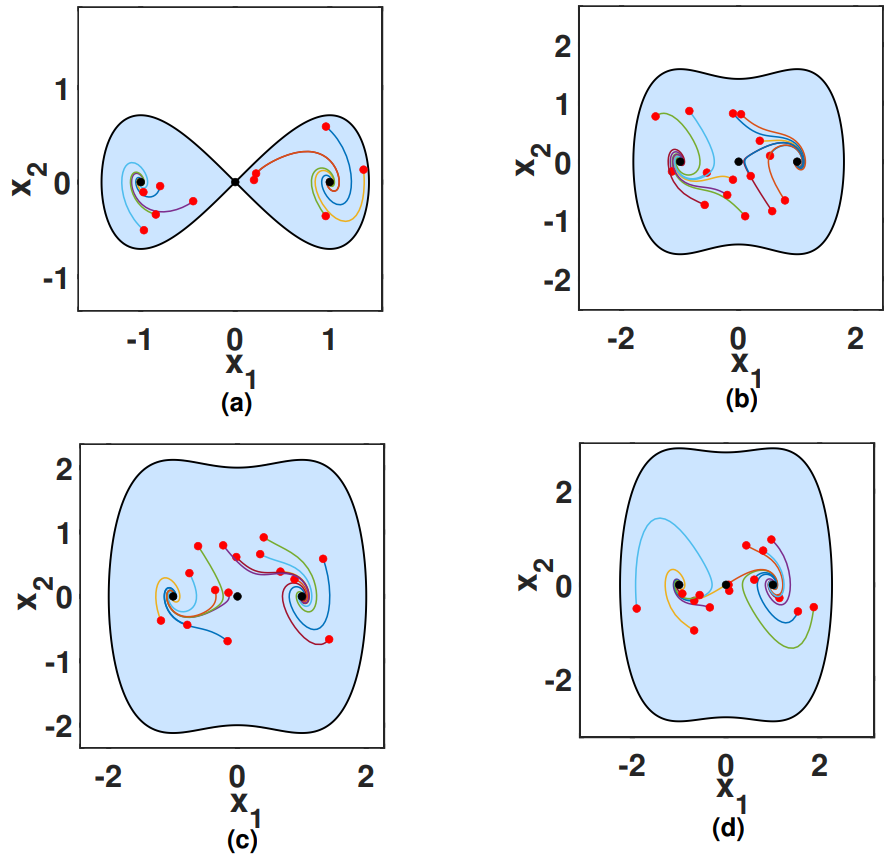} 
\caption{$\mathcal{D}_0$ regions (blue colored region) for four different values of $r$. Phase trajectories are shown for 10 different $\mathbf{x(0)} \in \mathcal{D}_0$. All the trajectories are converging to one of the equilibrium points. (a) For $r = 0$, (b) For $r =1$ (c) For $r = 2$, (d) For $r =4$.} 
\label{Diff_r_diff_U}
\end{center}
\end{figure}

So, the common BOA for the system \eqref{First example} is $\mathcal{D}_0 = \Omega \bigcap \mathcal{U}$. As $\Omega$ is the entire $\Re^2$ state space, so $\mathcal{D}_0 = \mathcal{U}$. The system is simulated to generate BOA for different values of $r$ as shown in Fig. \ref{Diff_r_diff_U}. Ten different initial conditions $\mathbf{x(0)} \in \mathcal{D}_0$ are used to simulate for computing phase trajectories. All the trajectories are converging to one of the equilibrium points. 

\subsection{Example 2: A system with $\Omega \subset \Re^2$}
The parameters for system \eqref{general nonlinear planar system} are chosen as follows: $b_1 = 4, b_2 = 3 > 0, b_3 = 1, b_4 = -3 < 0, q =0, m =1$ and $s =1$. 
\begin{equation} \label{2nd example}
\dot{\mathbf{x}} = 
\left[ \begin{array}{rr}
0 & 1\\
4 & 3
\end{array}\right]\mathbf{x} + 
\left[ \begin{array}{r}
0\\
 - x^3_1 - 3x^2_1x_2
\end{array}\right]
\end{equation}

The equilibrium points of system \eqref{2nd example} are $\mathbf{x^*_1}=(0,0)$, $\mathbf{x^*_2} = (2,0)$ and $\mathbf{x^*_3} = (-2,0)$. Let us now compute
\small
\begin{equation}
    \begin{split}
        \Omega & = \bigg\{(x_1,x_2) \in \Re^2 \bigg| 1 -x^2_1<0 \bigg\} \\
        \Omega_0 & = \bigg\{(x_1,x_2) \in \Re^2 \bigg| 1 -x^2_1=0\bigg\} \\
        \Omega_1 & = \bigg\{(x_1,x_2) \in \Re^2 \bigg| 1 -x^2_1>0 \bigg\}.
    \end{split}
\end{equation}
\normalsize
Let us consider the same $C$ (as the previous example); 
that encloses all $\mathbf{x^*_i};i=1,2,3$.
If we follow the identical steps as in Example 1, we get $I_C = +1$, and by similar analogy, it can be found that $\mathbf{x^*_1}$ is a saddle which is contained in $\Omega_1$, $\mathbf{x^*_2}$ and $\mathbf{x^*_3}$ are stable equilibrium points contained in $\Omega$.

\begin{figure}[h!]
\begin{center}
\includegraphics[width=7cm]{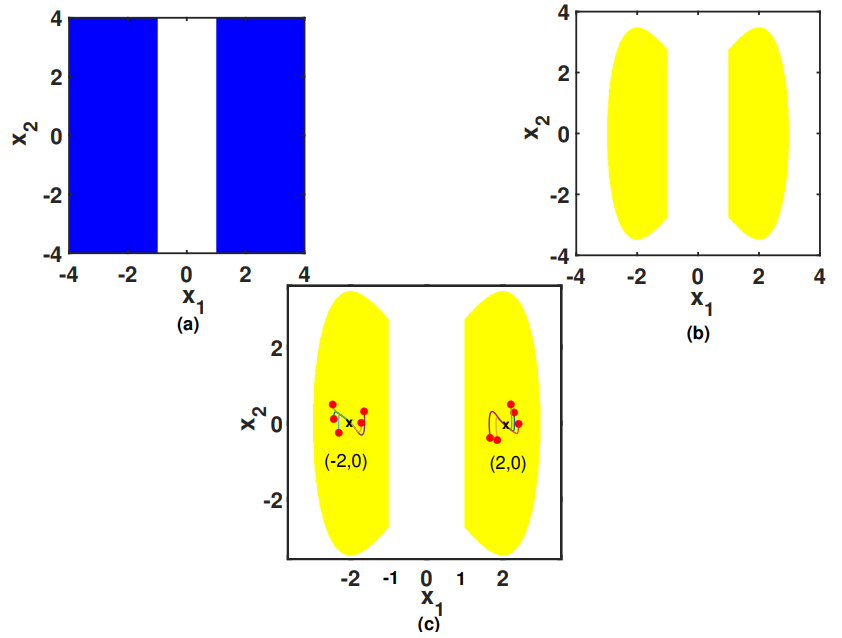} 
\caption{(a) $\Omega$ region (dark blue colored region), (b) $\mathcal{D}_0 = \Omega \bigcap \mathcal{U}$ (yellow colored region), (c) Phase trajectories originating from 10 different initial conditions $\mathbf{x(0)} \in \mathcal{D}_0$ and $\mathbf{x}(t)$ converge to one of the stable equilibrium points at $t \to \infty.$} 
\label{Example_2_NEW}
\end{center}
\end{figure}

According to Theorem \ref{Define D0 region}, $E(\mathbf{x})= \frac{1}{2}x^2_2 - 2x^2_1 + \frac{1}{4}x^4_1$ and $\frac{dE(\mathbf{x})}{dt} \leq 0, \forall t \geq 0$ in $\mathcal{U}$. So, the common BOA for the system \eqref{2nd example} is $\mathcal{D}_0 = \Omega \bigcap \mathcal{U}$. The system is simulated to generate $\Omega$ and $\mathcal{D}_0$ are shown in Fig. \ref{Example_2_NEW}. Ten different initial conditions $\mathbf{x(0)} \in \mathcal{D}_0$ are used to simulate for computing phase trajectories. All the trajectories are converging to one of the stable equilibrium points.

\subsection{Example 3: A system with no $\Omega$ region}
The parameters for system \eqref{general nonlinear planar system} are chosen as follows: $b_1 = 4, b_2 = 3 > 0, b_3 = 1, b_4 = 3 > 0, q =0, m =1$ and $s =1$. 
\begin{equation} \label{example 3}
\dot{\mathbf{x}} = 
\left[ \begin{array}{rr}
0 & 1\\
4 & 3
\end{array}\right]\mathbf{x} + 
\left[ \begin{array}{r}
0\\
 - x^3_1 + 3x^2_1x_2
\end{array}\right]
\end{equation}

For system \eqref{example 3}, $J^{[2]} = 3 + 3x^2_1 > 0, \forall t \geq 0, \forall \mathbf{x} \in \Re^2$. 
So, there is no $\Omega$ for this system \eqref{example 3}, hence there does not exist any $\mathcal{D}_0$.
\section{A network of communication between two agents: an application}
In multi-agent or network systems, the interactions among the components lead to the existence of multiple equilibrium points, even with the same set of inputs and/or initial conditions. Analyzing the stability of such systems is non-trivially challenging. To address this, we adopted the general procedure depicted in Section 3.

Consider a class of systems as follows:
\begin{equation} \label{NETWORKED SYSTEM}
    \begin{split}
        \dot{\mathbf{x}} & = -\Lambda \mathbf{x} + \pi W \Psi(\mathbf{x})
    \end{split}
\end{equation}
where;
   \begin{itemize}
   \item $ \mathbf{x} = [x_1 \hspace{0.2cm} x_2]^T $ be the state vector. $\mathbf{x_i}(t)$ represents the state of each node or agent at time instant $t$.
   \item \( \Delta = \text{diag}[\delta_1, \delta_2] \) represents the inertia or resistance of each agent to change their state. $\delta_i > 0, i=\{1,2\}$.
    \item \( W \) is the adjacency matrix representing how each agent’s state influences the others. It is an irreducible and hollow matrix, also assumed to be non-negative.
    \item \( \Psi(\mathbf{x}) = [\psi_1(x_1) \hspace{0.2cm} \psi_2(x_2) \)] is a nonlinear function representing the interactions between agents. Consider $\Psi(\mathbf{x}) = -\Psi(-\mathbf{x})$. Also, $\Psi(\mathbf{0})=\mathbf{0}$ and $\lim_{t \to \infty}\Psi(\mathbf{x}) = \pm 1$. Typical choices of $\Psi(\mathbf{x})$ are hyperbolic tangent function, Boltzmann functions $\Psi(x_i) = \frac{1-e^{-2x_i}}{1+e^{-2x_i}}$ etc.
     \item $\pi > 0:$ A scalar parameter.
   \end{itemize}

\subsection{Properties of the system \eqref{NETWORKED SYSTEM}}
System \eqref{NETWORKED SYSTEM} has only one equilibrium point $\mathbf{x^*_1}=(0,0)$ if $\pi <1$. At $\pi = 1$, the system \eqref{NETWORKED SYSTEM} undergoes a pitchfork bifurcation, creating $\mathbf{x^*_1}=(0,0)$, and the other two symmetric equilibrium points; i.e. $\mathbf{x^*_2}=(x^*_1,x^*_2)$ and $\mathbf{x^*_3}=(-x^*_1,-x^*_2)$. In \cite{fontan2017multiequilibria}, detailed bifurcation analysis has been done for $\pi > 1$. We want to demonstrate the efficacy of the use of 2-contraction analysis to find a common BOA if a system exhibits multiple stable equilibria. The system \eqref{NETWORKED SYSTEM} is not 1-contractive with respect to any norm. So we use the 2-contraction theory to study the behavior of the system. For system \eqref{NETWORKED SYSTEM}, we get $J^{[2]} = -(\delta_1 + \delta_2)$. So, $\mu_2(J^{[2]}) < 0, \forall t \geq 0, \forall \mathbf{x} \in \Re^2$. System \eqref{NETWORKED SYSTEM} has $\Omega = \{(x_1,x_2)|x_1 \in \Re, x_2 \in \Re\}$. As $\Omega$ is the entire $\Re^2$ space, by Lemma \ref{no periodic solution} no closed orbit or limit cycle can exist in $\Re^2$.

We provide Theorem \ref{LAST THEOREM} to compute $\mathcal{D}_0$ for the system \eqref{NETWORKED SYSTEM} adopting the algorithm given in Section \ref{section 3.3}. 

\begin{theorem} \label{LAST THEOREM}
Suppose, for system \eqref{NETWORKED SYSTEM} there exists a region $\Omega$ and also a simple closed curve $C$ such that $I_C(\mathbf{f}) = +1$. Then there exists a candidate energy function $E(\mathbf{x})$ given in \eqref{new energy function} 
\begin{equation} \label{new energy function}
    \begin{split}
        E(\mathbf{x}) & = \frac{1}{2}\mathbf{x}^T\mathbf{x}
    \end{split}
\end{equation}
with level set $\partial \mathcal{U} = \bigg\{(x_1,x_2) \in \Re^2 \bigg| E(\mathbf{x}) = r \bigg\}$ and 
\begin{equation} \label{new region hat{U}}
    \begin{split}
        \mathcal{U} & = \bigg \{(x_1,x_2) \in \Re^2 \bigg| E(\mathbf{x}) < r \bigg\}
    \end{split}
\end{equation}
such that $\frac{d E(\mathbf{x})}{dt} \leq 0$ in $\mathcal{U}$. Then common basin of attraction $\mathcal{D}_0$ is $\mathcal{D}_0 = \Omega \bigcap \mathcal{U}$, where $\Omega$ is the 2-contraction region for \eqref{NETWORKED SYSTEM}. The scalar $r$ can be selected to adjust the overall shape of $\mathcal{U}$.
\end{theorem} 

\proof
The first part of the proof, using index theory to analyze equilibrium behavior, aligns with the approach described in the proof of Theorem \ref{Define D0 region}. We find that $\mathbf{x^*_1}$ is saddle, $\mathbf{x^*_2}$ and $\mathbf{x^*_3}$ are stable equilibrium points. As already proved for system \eqref{NETWORKED SYSTEM}, $\Omega$ is the entire $\Re^2$ state space, so all the equilibrium points are contained in $\Omega$.

To compute the bounded region $\mathcal{U}$, let us evaluate
\begin{equation*}
    \begin{split}
   \frac{d E(\mathbf{x})}{dt} & = -\mathbf{x}^T \Lambda \mathbf{x} + \pi \mathbf{x}^T W \Psi(\mathbf{x}) 
    \end{split}
\end{equation*}
Now define, $\delta = \text{min}(\delta_1,\delta_2)>0$. In the second term $\pi \mathbf{x}^T W \Psi(\mathbf{x})$
\begin{itemize}
    \item $W$ is a hollow (zero diagonal) and non-negative adjacency matrix. This implies $\mathbf{x}^T W \mathbf{x} \geq 0$ for any $\mathbf{x}$, and hence $W \Psi(\mathbf{x})$ only depends on interactions between agents.
    \item Since $\Psi(\mathbf{x})$ is bounded ($\Psi(\mathbf{x}) \to \pm 1$ as $\mathbf{x} \to \infty$), the term $\pi \mathbf{x}^T W \Psi(\mathbf{x})$ is bounded for any $\mathbf{x}$.
\end{itemize}
Thus,
\begin{equation*}
    \begin{split}
   \parallel \pi \mathbf{x}^T W \Psi(\mathbf{x}) \parallel \leq \pi \parallel \mathbf{x} \parallel \parallel W \Psi(\mathbf{x}) \parallel \leq C_1 \parallel \mathbf{x} \parallel     
    \end{split}
\end{equation*}
where $C_1$ is a constant depending on $\pi$, $W$, and the boundedness of $\Psi(\mathbf{x})$. Now we get,
\begin{equation*}
    \begin{split}
  \frac{d E(\mathbf{x})}{dt} & \leq -\delta \parallel \mathbf{x} \parallel^2 + C_1 \parallel \mathbf{x} \parallel \\
\Rightarrow \frac{d E(\mathbf{x})}{dt} & \leq -\delta \parallel \mathbf{x} \parallel \bigg( \parallel \mathbf{x} \parallel - \frac{C_1}{\delta} \bigg) 
    \end{split}
\end{equation*}
To comment on the boundedness of trajectories, consider two cases:
\begin{itemize}
    \item When $\parallel \mathbf{x} \parallel \geq \frac{C_1}{\delta}$ then $\frac{d E(\mathbf{x})}{dt} \leq 0$.
 \item When $\parallel \mathbf{x} \parallel \leq \frac{C_1}{\delta}$ then $\frac{d E(\mathbf{x})}{dt} \leq 0$.
\end{itemize}
Thus, the system is globally bounded, with $\frac{C_1}{\delta}$ serving as an upper bound on $\parallel \mathbf{x} \parallel$. That suggests $r$ can be chosen as large as possible. Finally $\mathcal{D}_0 = \Omega \bigcap\mathcal{U} = \mathcal{U}$ as $\Omega$ is the entire $\Re^2$ state space. $\blacksquare$

\subsection{Nonlinear opinion dynamics model}
\begin{figure}[h!]
\begin{center}
\includegraphics[width=4cm]{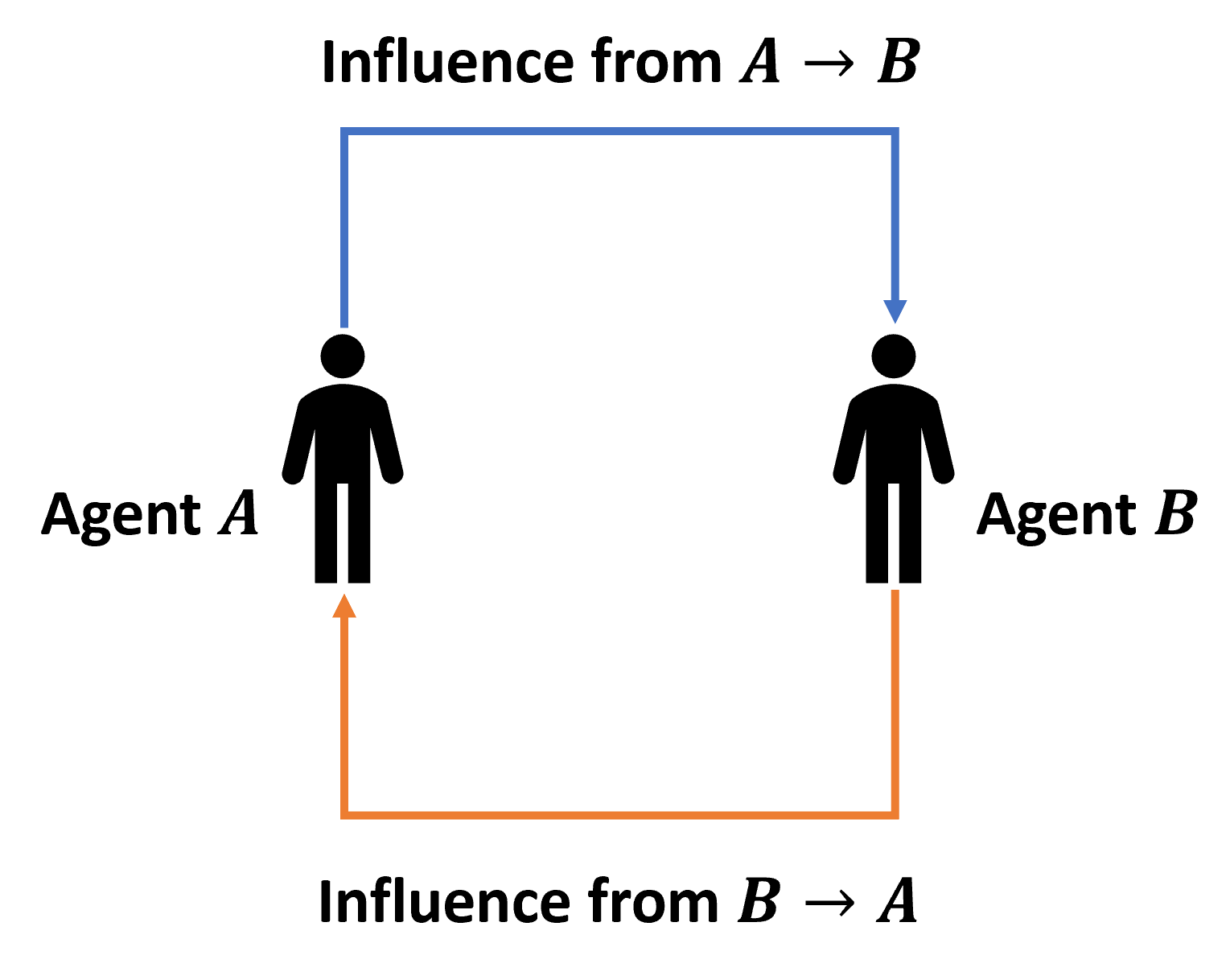} 
\caption{Pictorial representation of an opinion dynamics model between two agents.}
\label{2 Agents}
\end{center}
\end{figure}

Consider a model for nonlinear opinion dynamics between two agents as depicted in Fig. \ref{2 Agents}.

The general networked system model \eqref{NETWORKED SYSTEM} can be used in this context. Now, $x_i(t)$ represents the opinion of $i^{th}$ agent at time $t$. The $-\delta_i x_i$ represents the ``forgetting term". $\pi$ represents a community social effort \cite{fontan2017multiequilibria}. For the analysis purpose we consider; $\pi = 2, \delta_1 = 0.2, \delta_2 = 0.4, W = 
\left[ \begin{array}{rr}
0 & 0.5\\
0.7 & 0
\end{array}\right], \Psi(x_i)= \text{tanh}(x_i)$. We consider no self-influence in this example; i.e. $w_{11} = 0, w_{22} = 0$. This parameter set up gives $\mathbf{x^*_1}=(0,0), \mathbf{x^*_2}=(-4.99,-3.499), \mathbf{x^*_3} = (4.99,3.499)$. Using index theory we find out $\mathbf{x^*_1}$ is saddle and $\mathbf{x^*_i},i=2,3$ are stable equilibrium points. Also, as the system is globally 2-contractive, there does not exist any cyclic pattern or periodic behavior of opinions between the agents. As proved earlier the system trajectories are bounded.

\begin{figure}[h!]
\begin{center}
\includegraphics[width=8cm]{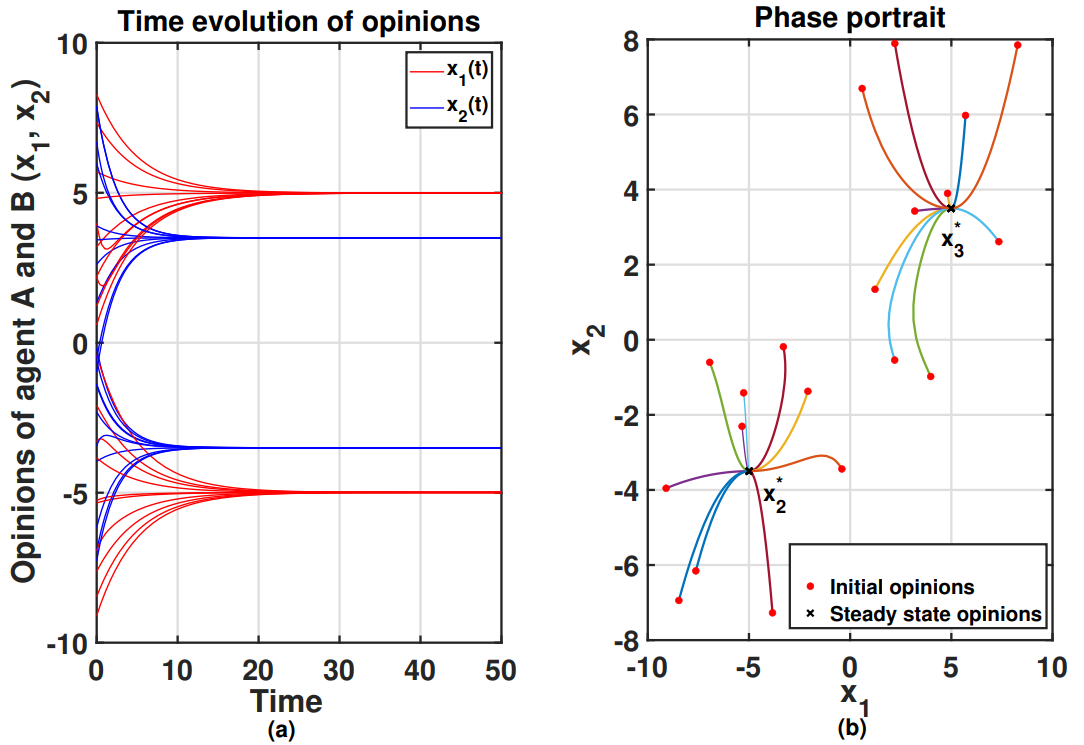} 
\caption{Evolution of opinions of agents A and B are shown. (a) Polarization of opinions between two agents, (b) Phase trajectories.}
\label{Opinion_dynamics}
\end{center}
\end{figure}

According to Theorem \ref{LAST THEOREM}, we can select $\mathcal{U} = \bigg\{(x_1,x_2) \in \Re^2 \bigg| \frac{1}{2}x^2_1 + \frac{1}{2}x^2_2 < r \bigg\}$, where $E(\mathbf{x}) = \frac{1}{2}x^2_1+\frac{1}{2}x^2_2$ and $\frac{d E}{dt} \leq 0, \forall t \geq 0$ in $\mathcal{U}$. Finally, $\mathcal{D}_0 = \Omega \bigcap \mathcal{U} = \mathcal{U}$ as $\Omega$ is the entire $\Re^2$.
Now, if the initial opinion $\mathbf{x(0)} \in \mathcal{D}_0$ then $\mathbf{x}(t) \to \mathbf{x^*_i}, i=\{1,2,3\}$ as $t \to \infty$.  On the other hand if $\mathbf{x(0)} \in \mathcal{D}_0-\mathcal{M}_s$, where $\mathcal{M}_s$ is the stable manifold of saddle equilibrium point $\mathbf{x^*_1}$, then $\mathbf{x}(t)\to \mathbf{x^*_i}, i=\{2,3\}$. So it is observed that despite having polarization in the dynamics, each agent can achieve a stable opinion at a steady state. In fact, careful consideration of initial opinions and choice of parameters $\{\pi, \delta_i,w_{ij}\}$ may lead to a consensus of opinions between two agents. As for example; if $\pi > 1$ and $w_{12}=w_{21}, \delta_1 =  \delta_2$ then $\mathbf{x^*_2} =(x^*,x^*), \mathbf{x^*_3}=(-x^*,-x^*)$. Fig. \ref{Opinion_dynamics} depicts the evolution of opinions of agents A and B where twenty different initial opinions (initial conditions) are chosen randomly. It is seen that that all the opinions converge to one of the steady state as $t \to \infty$.
\section{Conclusions}
An analytical procedure is proposed to find a common basin of attraction for a general nonlinear planar system that exhibits multiple isolated stable equilibrium points. Further, the procedure is adopted to characterize the common basin of attraction. With the help of Poincaré index theory, we have effectively gained insights into the behavior of equilibrium points without relying on local analysis. Leveraging the results of 2-contraction and Poincaré index theory, a methodology is proposed to compute a common basin of attraction. Further, this methodology has been applied to different benchmark examples including an opinion dynamics problem to study the global behavior of the system. Numerical simulations are employed in every example to validate the theoretical results.

The immediate extension is to apply such an analytical approach to higher-order nonlinear systems $(n > 2)$. The proposed region of interest $\mathcal{D}_0$ is not unique, nor claimed to be the best choice for a common basin of attraction. The characterization and computation of such regions can be improved by considering different energy-like functions. 

\section*{Appendix A}
\subsection*{Calculation of $I_C$}
The parametrized form of $C$ can be written as:

\begin{equation*}
    \begin{split}
        C & = \bigg\{ \bigg(x_1(\theta),x_2(\theta)\bigg)\bigg|x_1(\theta)= 4\text{cos}(\theta), x_2(\theta) = 4\text{sin}(\theta); \\
        & \theta\in [0,2\pi)  \bigg\}
    \end{split}
\end{equation*}
The vector field $\mathbf{f(x)}$ along $C$: $\mathbf{f}(\theta) =  \bigg[f_1(\theta) \hspace{0.1cm} 
f_2(\theta) \bigg]^T$. Where, 
\begin{equation*}
    \begin{split}
f_1(\theta) & = 4 \text{sin}(\theta) \\
f_2(\theta) & = -4 \text{sin}(\theta) + 4 \text{cos}(\theta) - 64 \text{cos}^3(\theta) 
    \end{split}
\end{equation*}
To compute the \textbf{total change in angle} ($\Delta \Phi$) of $\mathbf{f}$ as $\theta$ goes from $0$ to $2\pi$, follow these steps:
\begin{itemize}
    \item The angle $\Phi(\theta)$ of the vector $\mathbf{f}(\theta)$ is given by:
\begin{equation*}
    \begin{split}
\Phi(\theta) = \arctan\left( \frac{f_2(\theta)}{f_1(\theta)} \right)       
    \end{split}
\end{equation*}
\item Let us track the Change in $\Phi(\theta)$: We tabulate the angle at key points and compute the incremental changes (refer to Table 2):
\begin{table}[h!] \label{Table 2}
\centering
\begin{tabular}{|c|c|c|c|c|}
\hline
$\theta$ & $f_1(\theta)$ & $f_2(\theta)$ & $\Phi(\theta)$ & $\Delta \Phi$  \\
\hline
$0$ & $0$ & $-60$ & $270^\circ$ & - \\
\hline
$\frac{\pi}{2}$ & $4$ & $-4$ & $315^\circ$ & $+45^\circ$ \\
\hline
$\pi$ & $0$ & $60$ & $90^\circ$ & $+135^\circ$ \\
\hline
$\frac{3\pi}{2}$ & $-4$ & $4$ & $135^\circ$ & $+45^\circ$ \\
\hline
$2\pi$ & $0$ & $-60$ & $270^\circ$ & $+135^\circ$ \\
\hline
\end{tabular}
\caption{Angle $\Phi(\theta)$ and its incremental changes.}
\end{table}
\normalsize
\item Total change in $\Phi(\theta)$:
\begin{equation*}
    \begin{split}
\Delta \Phi = 45^\circ + 135^\circ + 45^\circ + 135^\circ = 360^\circ \hspace{0.1cm} (\text{or } 2\pi \text{ radians})        
    \end{split}
\end{equation*}
\item Calculating $I_C$ using \eqref{POINCARE INDEX FORMULA}, we get
\begin{equation*}
    \begin{split}
  I_C & = \frac{\Delta \Phi}{2\pi} = \frac{2\pi}{2\pi} = +1.  
    \end{split}
\end{equation*}
\end{itemize}
\section*{Declarations}

\textbf{Conflict of interest :}
The authors declare that they have shared no conflict of interest.

\textbf{Funding :}
No funding has been received to assist with this research work.

\textbf{Data Availability :}
The data that supports the findings of this
the study is available from the corresponding author upon reasonable request.

\section*{Acknowledgment}
The author wants to acknowledge the anonymous reviewers for their valuable comments in the first review. The technical rigour of the comments helps us to improve the manuscript a lot

\bibliographystyle{asmems4}
\bibliography{asme2e}

\begin{thebibliography}{10}

\bibitem{xiong2018social}
Xiong, F., Wang, X., Pan, S., Yang, H., Wang, H., and Zhang, C., 2018,
\newblock ``Social recommendation with evolutionary opinion dynamics,''
\newblock {\em IEEE Transactions on Systems, Man, and Cybernetics: Systems, {\bf 50}}(10), pp.~3804--3816.

\bibitem{proskurnikov2015opinion}
Proskurnikov, A.~V., Matveev, A.~S., and Cao, M., 2015,
\newblock ``Opinion dynamics in social networks with hostile camps: Consensus vs. polarization,''
\newblock {\em IEEE Transactions on Automatic Control, {\bf 61}}(6), pp.~1524--1536.

\bibitem{doi:10.1137/130913250}
Jia, P., MirTabatabaei, A., Friedkin, N.~E., and Bullo, F., 2015,
\newblock ``Opinion dynamics and the evolution of social power in influence networks,''
\newblock {\em SIAM Review, {\bf 57}}(3), pp.~367--397.

\bibitem{hesaaraki2013global}
Hesaaraki, M., and Sabzevari, M., 2013,
\newblock ``Global properties for an hiv-1 infection model including an eclipse stage of infected cells and saturation infection.,''
\newblock {\em Differential Equations \& Control Processes, {\bf 2013}}(4).

\bibitem{intissar2020application}
Intissar, A., 2020,
\newblock ``Application of the criterion of li-wang to a five dimensional epidemic model of covid-19. part i,''
\newblock {\em arXiv preprint arXiv:2007.07815}.

\bibitem{RON_OFIR}
Ofir, R., Ovseevich, A., and Margaliot, M., 2024,
\newblock ``Contraction and k-contraction in lurie systems with applications to networked systems,''
\newblock {\em Automatica, {\bf 159}}, p.~111341.

\bibitem{9782543}
Ofir, R., Margaliot, M., Levron, Y., and Slotine, J.-J., 2022,
\newblock ``A sufficient condition for $k$-contraction of the series connection of two systems,''
\newblock {\em IEEE Transactions on Automatic Control, {\bf 67}}(9), pp.~4994--5001.

\bibitem{lohmiller1998contraction}
Lohmiller, W., and Slotine, J.-J.~E., 1998,
\newblock ``On contraction analysis for non-linear systems,''
\newblock {\em Automatica, {\bf 34}}(6), pp.~683--696.

\bibitem{jouffroy2005some}
Jouffroy, J., 2005,
\newblock ``Some ancestors of contraction analysis,''
\newblock In Proceedings of the 44th IEEE Conference on Decision and Control, IEEE, pp.~5450--5455.

\bibitem{aminzare2014contraction}
Aminzare, Z., and Sontag, E.~D., 2014,
\newblock ``Contraction methods for nonlinear systems: A brief introduction and some open problems,''
\newblock In 53rd IEEE Conference on Decision and Control, IEEE, pp.~3835--3847.

\bibitem{jouffroy2005some123}
Jouffroy, J., and Fossen, T.~I., 2010,
\newblock ``A tutorial on incremental stability analysis using contraction theory,''
\newblock {\em Modeling, Identification and Control, {\bf 31}}(3), p.~93.

\bibitem{aminzare2014synchronization}
Aminzare, Z., and Sontag, E.~D., 2014,
\newblock ``Synchronization of diffusively-connected nonlinear systems: Results based on contractions with respect to general norms,''
\newblock {\em IEEE transactions on network science and engineering, {\bf 1}}(2), pp.~91--106.

\bibitem{partialcontraction}
Wang, W., and Slotine, J.-J.~E., 2005,
\newblock ``On partial contraction analysis for coupled nonlinear oscillators,''
\newblock {\em Biological cybernetics, {\bf 92}}(1), pp.~38--53.

\bibitem{sharma2008design}
Sharma, B.~B., and Kar, I.~N., 2008,
\newblock ``Design of asymptotically convergent frequency estimator using contraction theory,''
\newblock {\em IEEE Transactions on Automatic Control, {\bf 53}}(8), pp.~1932--1937.

\bibitem{forni2013differential}
Forni, F., and Sepulchre, R., 2013,
\newblock ``A differential lyapunov framework for contraction analysis,''
\newblock {\em IEEE Transactions on Automatic Control, {\bf 59}}(3), pp.~614--628.

\bibitem{tsukamoto2021contraction}
Tsukamoto, H., Chung, S.-J., and Slotine, J.-J.~E., 2021,
\newblock ``Contraction theory for nonlinear stability analysis and learning-based control: A tutorial overview,''
\newblock {\em Annual Reviews in Control, {\bf 52}}, pp.~135--169.

\bibitem{simpson2014contraction}
Simpson-Porco, J.~W., and Bullo, F., 2014,
\newblock ``Contraction theory on riemannian manifolds,''
\newblock {\em Systems \& Control Letters, {\bf 65}}, pp.~74--80.

\bibitem{chaffey2018control}
Chaffey, T.~L., and Manchester, I.~R., 2018,
\newblock ``Control contraction metrics on finsler manifolds,''
\newblock In 2018 Annual American Control Conference (ACC), IEEE, pp.~3626--3633.

\bibitem{giesl2022review}
Giesl, P., Hafstein, S., and Kawan, C., 2022,
\newblock ``Review on contraction analysis and computation of contraction metrics,''
\newblock {\em arXiv preprint arXiv:2203.01367}.

\bibitem{russo2012contraction}
Russo, G., Di~Bernardo, M., and Sontag, E.~D., 2012,
\newblock ``A contraction approach to the hierarchical analysis and design of networked systems,''
\newblock {\em IEEE Transactions on Automatic Control, {\bf 58}}(5), pp.~1328--1331.

\bibitem{WU2022110048}
Wu, C., Kanevskiy, I., and Margaliot, M., 2022,
\newblock ``k-contraction: Theory and applications,''
\newblock {\em Automatica, {\bf 136}}, p.~110048.

\bibitem{harzheim2005ordered}
Harzheim, E., 2005,
\newblock {\em Ordered sets}, Vol.~7
\newblock Springer Science \& Business Media.

\bibitem{deutsch1975matrix}
Deutsch, E., 1975,
\newblock ``On matrix norms and logarithmic norms,''
\newblock {\em Numerische Mathematik, {\bf 24}}, pp.~49--51.

\bibitem{martin1974logarithmic}
Martin~Jr, R.~H., 1974,
\newblock ``Logarithmic norms and projections applied to linear differential systems,''
\newblock {\em Journal of Mathematical Analysis and Applications, {\bf 45}}(2), pp.~432--454.

\bibitem{vidyasagar2002nonlinear}
Vidyasagar, M., 2002,
\newblock {\em Nonlinear systems analysis}
\newblock SIAM.

\bibitem{li1995ra}
Li, M.~Y., and Muldowney, J.~S., 1995,
\newblock ``On ra smith's autonomous convergence theorem,''
\newblock {\em The Rocky Mountain Journal of Mathematics}, pp.~365--379.

\bibitem{meiss2007differential}
Meiss, J.~D., 2007,
\newblock {\em Differential dynamical systems}
\newblock SIAM.

\bibitem{strogatz2001nonlinear}
Strogatz, S.~H., 2001,
\newblock ``Nonlinear dynamics and chaos: with applications to physics, biology, chemistry, and engineering (studies in nonlinearity),''
\newblock {\em Nonlinear Dynamics and Chaos: With Applications to Physics, Biology, Chemistry, and Engineering (Studies in Nonlinearity)}.

\bibitem{rao1995mechanical}
Rao, S.~S., and Yap, F.~F., 1995,
\newblock {\em Mechanical vibrations}, Vol.~4
\newblock Addison-Wesley New York.

\bibitem{zhu2013stochastic}
Zhu, W., Cai, G., and Hu, R., 2013,
\newblock ``Stochastic analysis of dynamical system with double-well potential,''
\newblock {\em International Journal of Dynamics and Control, {\bf 1}}, pp.~12--19.

\bibitem{fontan2017multiequilibria}
Fontan, A., and Altafini, C., 2017,
\newblock ``Multiequilibria analysis for a class of collective decision-making networked systems,''
\newblock {\em IEEE Transactions on Control of Network Systems, {\bf 5}}(4), pp.~1931--1940.

\end{thebibliography}
\end{document}